\pdfoutput=1

\documentclass[twocolumn,english]{revtex4}
\usepackage{times,amssymb,amsmath,graphicx}
\usepackage[bookmarks=false,pdffitwindow=false,pdfstartview={FitH}]{hyperref}
\usepackage[T1]{fontenc}
\usepackage{babel,multirow}
\usepackage{color}

\begin{document}


\title{
  Josephson effect in graphene Corbino disks
}

\author{Adam Rycerz\footnote{Correspondence: 
  \href{mailto:rycerz@th.if.uj.edu.pl}{rycerz@th.if.uj.edu.pl}.}}
\affiliation{Institute for Theoretical Physics,
  Jagiellonian University, \L{}ojasiewicza 11, PL--30348 Krak\'{o}w, Poland}

\date{September 15, 2026}

\begin{abstract}
Peculiar features of the Josephson effect in graphene were described 
theoretically by Titov and Beenakker 
[\href{https://doi.org/10.1103/PhysRevB.74.041401}{%
Phys.\ Rev.\ B {\bf 74}, 041401(R) (2006)}],  who solved the
Dirac-Bogoliubov-de-Gennes equation for
a~superconductor-graphene-superconductor junction with rectangular geometry.
Here, we adopt the analysis for graphene Corbino disks, finding out that ---
for the outer to inner radii ratio $r_2/r_1\gtrsim{}5$ --- such systems
may demonstrate, when varying the electrochemical potential and the spatial
profile of the electrostatic barrier, crossover from standard Josephson
tunneling (SJT), via graphene-specific multimode Dirac-Josephson tunneling
(MDJT), towards the ballistic Josephson effect (BJE). 
Signatures of SJT appear only near the Dirac point 
when the barrier shape is close to rectangular, MDJT appears
in the tripolar range and is very robust against varying the barrier shape,
and BJE is restored in the unipolar range when smoothing the barrier shape.
A~comparison with the results of a~numerical simulation of quantum transport
on the honeycomb lattice is also given. 
\end{abstract}

\maketitle


\section{Introduction}

The disk-shaped device, consists of an annular conductor, attached to inner
and outer highly conducting electrodes covering the inner and outer circular
periphery, was considered by Boltzmann in 1886 \cite{Bol86} and by several
authors the in early 1900s \cite{Cor11,Ada15} to measure the magnetoresistance
without generating the Hall voltage, making a~significant step towards
understanding the nature of charge transport in ordinary solids \cite{Gal91}. 
Later, an interest in --- as commonly called --- Corbino geometry has
reappeared after the fabrication of
GaAs/AlGaAs heterostructures \cite{Kir94,Sou98,Man96,dAm13}, and the 
discovery of high-temperature superconductivity \cite{Ryc99}, as edge-free
devices provide valuable insights into the system dynamics. 

\begin{figure}[b]
\includegraphics[width=\linewidth]{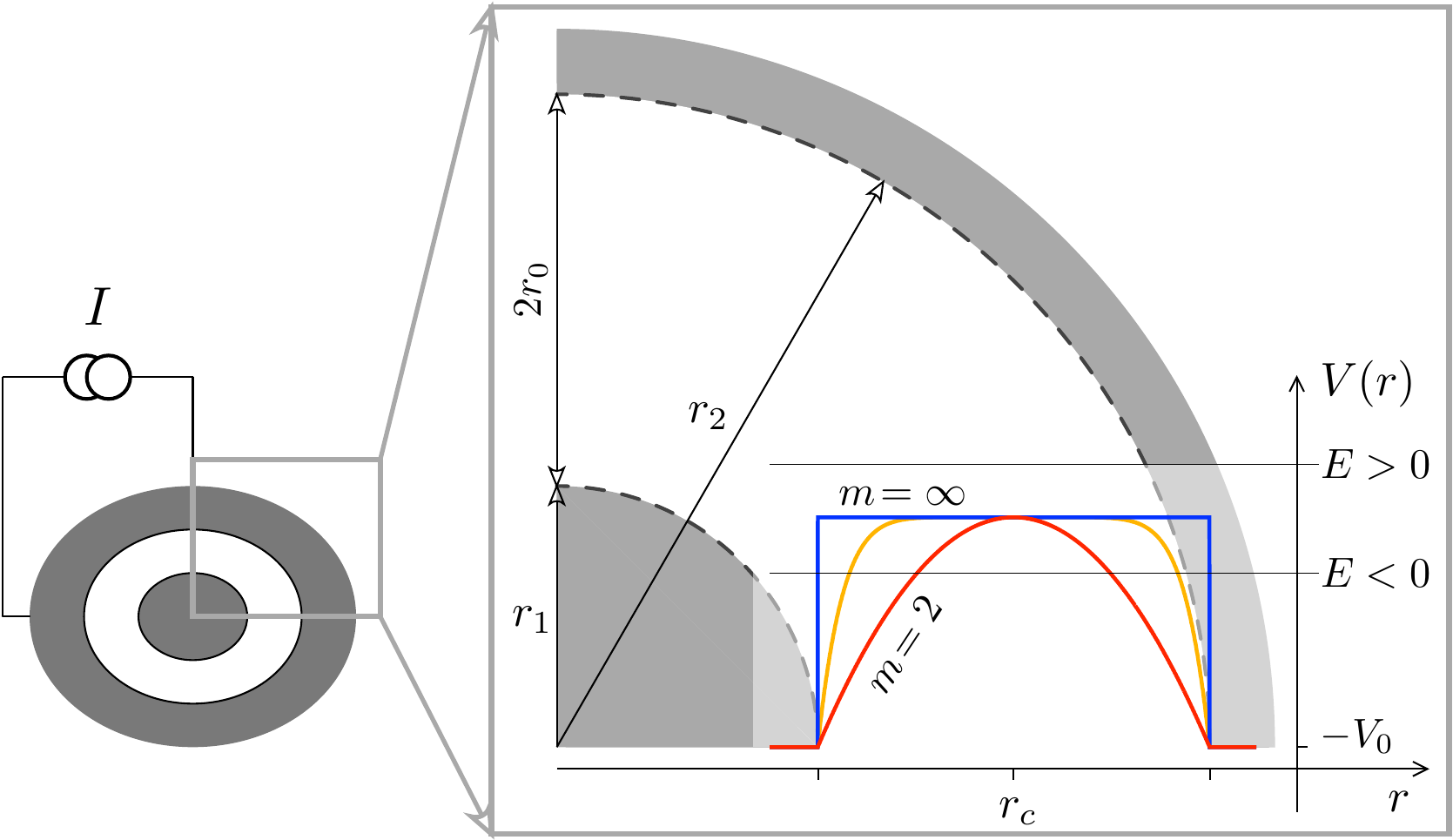}
\caption{ \label{diskmpot} 
  Left: Schematic of a~graphene disk with inner radius $r_1$ and outer radius
  $r_2$, contacted by two circular superconducting electrodes (dark areas). 
  A current source drives a~dissipationless supercurrent through the annular
  region (white). A separate gate electrode (not shown) allows one to tune
  the carrier concentration around the neutrality point.
  Right: Electrostatic potential profiles given by Eq.\ (\ref{vrmpot})
  with $m=2$, $8$ and $m=\infty$ (i.e., the rectangular barrier).
  The Fermi energy $E$ is defined with respect to the top of a barrier.
  $E > 0$ corresponds to unipolar n-n-n doping in the device; for $E<0$,
  circular n-p-n (tripolar) structure is formed. Arcs (dashed lines)
  mark the interfaces between the disk area [$r_1<r<r_2$] and contact
  regions [$r<r_1$ or $r>r_2$]. 
}
\end{figure}

In the context of graphene, the edge-free Corbino geometry is often considered
\cite{Kat20,Che06,Ryc09,Ryc10,Kha13,Pet14,Abd17,Ryc21a,Zen19,Sus20,Kam21,%
Yer21}, 
mainly because, in such a~geometry, magnetotransport at high fields is
unaffected by edge states, allowing one to probe the bulk transport
properties \cite{Zen19,Sus20,Kam21,Yer21}.
Although the Josephson effect in graphene has attracted considerable
attention
\cite{Tit06,Mog06,Hag10,Ali11,Gua15,Cal15,Eng16,Nan17,Woj18,Abd18,Hua22,%
Zha23,Ban24,Jan25,Ryc26a},
the Corbino-Josephson setup (i.e., the disk contacted to superconducting
electrodes, see Fig.\ \ref{diskmpot}), earlier studied both in
superconductor-insulator-superconductor \cite{Cri00} and
superconductor-normal metal–superconductor \cite{Had03} variants,
and recently constructed on the surface of topological insulators
\cite{Zha22}, 
has been discussed only marginally in the context of graphene \cite{Abd18}. 

In this paper, we consider the Corbino geometry
superconductor-graphene-superconductor junction, focusing on the wide-disk
case (the outer radius $r_2\gg{}r_1$---the inner radius). 
In such a~case, in the Dirac point, the transport is governed by two equivalent
(and fourfold-degenerate) modes, corresponding to total angular momentum
quantum numbers $j=\pm{}1/2$, with the transmission probabilities
$T_{\pm{}1/2}\propto{}r_1/r_2\ll{}1$ \cite{Ryc09,Ryc10}.
Therefore, a~standard Josephson tunneling (SJT), characterized by a~sinusoidal
current-phase relation, is expected near the Dirac
point, away from which graphene-specific multimode Dirac-Josephson tunneling
(MDJT) \cite{Tit06,Ryc26a} may appear.
In addition, the radial electrostatic potential profile is tuned, from 
a~rectangular to a~parabolic one, making the device capable of
showing also the ballistic Josephson effect (BJE), building analogy with
the sub-Sharvin to ballistic crossover predicted for normal metallic
electrodes \cite{Ryc21b,Ryc22}.

As our discussion is limited to zero temperature, hydrodynamic effects,
which may alter the system characteristics starting from few-kelvin 
temperatures \cite{Tom14,Kum22,Lev22,Vij25}, are beyond the scope of
this work.

The paper is organized as follows.
In Sec.\ \ref{modmet} we present the details of our numerical approach. 
The results for the Corbino-Josephson setup in graphene with 
rectangular potential barrier are summarized in Sec.\ \ref{recbar}. 
The central results of the paper, concerning the critical current and
skewness of the current-phase relation for smooth potentials, are presented
in Sec.\ \ref{smopot}. 
A comparison with the tight-binding simulation of quantum
transport is provided in Sec.\ \ref{tbasim}. 
The concluding remarks are given in Sec.\ \ref{conclu}.

\section{Model and methods}
\label{modmet}

\subsection{Dirac Hamiltonian for the normal state}
In typical graphene nanosystems, boundary effects may strongly affect
transport properties of the system, not only near the Dirac point, where
the transport is governed by one (or a~few) discrete modes \cite{Two06,Mia07}
but also away from the Dirac point \cite{Ryc25}, albeit to a~smaller degree.
In the case of edge-free Corbino geometry, the role of the boundary effects
is eliminated; instead, the role of interfaces separating the sample and
the leads becomes essential.
In particular, for normal metallic leads, significant particle-hole asymmetry
of the conductance spectrum was observed \cite{Pet14,Kam21,Lai16}; namely,
the conductance for the chemical potential $\mu<0$ (defined with respect to
the charge-neutrality point) is noticeably suppressed 
compared with the $\mu>0$ range.  This observation is interpreted in terms
of additional, contact resistance which is amplified in case of tripolar
n-p-n doping, compared to unipolar n-n-n doping, as the the circular p-n
junctions introduce additional backscattering of electrons in the former case. 
In Ref.\ \cite{Ryc21b}, a~simple model capable of providing a~qualitatively
correct description of the above-mentioned asymmetry, is put forward.
(Such a feature is also correctly reproduced by a~model assuming the
trapezoidal potential barrier \cite{Par21}, which allows a fully analytical
treatment; but this approach produces an artificial conductance maximum
near $\mu=0$.) 

Our approach starts from the Dirac Hamiltonian for low-energy excitations
in graphene in the normal state, for $K$ valley, 
\begin{equation}
\label{ham0dir}
  \mathcal{H}_0=v_F\,\mbox{\boldmath$p$}\cdot\mbox{\boldmath$\sigma$}
  + V({\bf r}), 
\end{equation}
with
$v_F=\sqrt{3}\,t_0a_0/(2\hbar)\approx{}10^6\,$m$/$s the
energy-independent Fermi velocity ($t_0=2.7\,$eV is 
the nearest-neighbor hopping integral and $a_0=0.246$ is the lattice
parameter), 
and $\mbox{\boldmath$p$}=(p_x,p_y)$ the in-plane momentum operator 
(with $p_j=-i\hbar{}\partial_j$), $\mbox{\boldmath$\sigma$}=(\sigma_x,\sigma_y)$
(with $\sigma_j$ being the Pauli matrices).
For the forthcoming numerical calculations, we set (in the physical units)
$\hbar{}v_F=0.575214\,$eV$\cdot$nm. 
The electrostatic potential energy in Eq.\ (\ref{ham0dir}) depends --- in
the polar coordinates $(r,\varphi)$ --- only on $r$, and is given by 
\begin{equation}
\label{vrmpot}
  V({\bf r}) = V(r) = -V_0\times
  \begin{cases}
    \,1  &  \text{if }\ |r-r_c| > r_0, \\
    \,\frac{|r-r_c|^m}{r_0^m}  &  \text{if }\ |r-r_c| \leqslant r_0, 
  \end{cases}
\end{equation}
where we have defined $r_c=(r_1+r_2)/2$ and $r_0=(r_2-r_1)/2$.
In particular, the limit of $m\rightarrow{}\infty$ corresponds to the
rectangular barrier (with a~cylindrical symmetry); any finite $m\geqslant{}2$
defines
a~smooth potential barrier, interpolating between the parabolic ($m=2$)
and rectangular ($m=\infty$) shape.
In principle, barrier smoothing can be regarded 
as a~feature of a~self consistent solution originating from the diffusion
of carriers; we expect this feature to strongly depend on the experimental
details, with graphene-on-hBN devices \cite{Zen19} showing rectangular,
rather than smooth, profiles.

\subsection{The Dirac-Bogoliubov-De-Gennes equation}
In the presence of superconducting leads (the $r<r_1$ and $r>r_2$ regions)
we employ the Dirac-Bogoliubov-de~Gennes equation \cite{Tit06,Mog06}
\begin{equation}
\label{dbdgeq}
\begin{pmatrix}
 \mathcal{H}_0-\mu & \Delta \\
 \Delta^\star & \mu-\overline{\mathcal{H}}_0
\end{pmatrix}
\begin{pmatrix}
 \Psi_e \\
 \Psi_h
\end{pmatrix} 
=
\varepsilon\begin{pmatrix}
 \Psi_e \\
 \Psi_h
\end{pmatrix}.  
\end{equation}
Here, $\Psi_e$ and $\Psi_h$ are the electron and hole wave functions, 
$\varepsilon>0$ is the excitation energy, and 
$\mu=E$ --- the Fermi energy --- since the $T=0$ case is considered. 
In the absence of a magnetic field, the Hamiltonian is time-reversal
invariant, 
$\overline{\mathcal{H}}_0
=\mathcal{T}\mathcal{H}_0\mathcal{T}^{-1}=\mathcal{H}_0$,
with $\mathcal{T}$ the time-reversal operator \cite{Bee06}. 

The complex pair potential $\Delta$ in Eq.\ (\ref{dbdgeq}) depends only the
radial coordinate ($r$), and is truncated by adopting the step-function
model for the two interfaces between the normal region and superconductors 
at $r=r_1$ and $r=r_2$, namely
\begin{equation}
  \Delta({\bf r}) = \Delta(r) = 
  \begin{cases}
    \,\Delta_0e^{i\theta/2}  &  \text{if }\ r < r_1, \\
    \,0  &  \text{if }\ r_1\leqslant{}r\leqslant{}r_2, \\
    \,\Delta_0e^{-i\theta/2}  &  \text{if }\ r > r_2,   
  \end{cases}
\end{equation}
with the bulk superconducting gap $\Delta_0$ and the phase difference
between the superconductors $\theta$.
(A~possible adaptation --- for the case of S-g-S junctions --- of the
self-consistent description of $\Delta({\bf r})$, earlier developed for
standard (non-relativistic) Josephson junctions \cite{Zar99,Gum07},
is beyond the scope of this work.) 

As shown in Refs.\ \cite{Tit06,Mog06}, by analyzing the spectrum of
Andreev states for $\varepsilon<\Delta_0$ for the short-junction limit, 
i.e., $r_2-r_1\ll{}\xi_0$ (with the superconducting coherence length
$\xi_0=\hbar{}v_F/\Delta_0$; for instance, $\xi_0\approx{}550\,$nm
for superconducting electrodes made with molybdenum rhenium \cite{Nan17}),
the Josephson current can be written as
\begin{equation}
\label{ijoth}
  I(\theta)=\frac{e\Delta_0}{\hbar}\sum_{j}
  \frac{T_j\sin\theta}{\sqrt{1-T_j\sin^2(\theta/2)}},  
\end{equation}
while the normal-state resistance is given by
\begin{equation}
\label{rnlan}
  R_N^{-1}=\frac{4e^2}{h}\sum_{j}T_j.
\end{equation}
In effect, both quantities are determined by the transmission probabilities
$T_j$ characterizing a~graphene sample between two electrodes in the normal
state ($\Delta_0=0$). 
Eqs.\ (\ref{ijoth}) and (\ref{rnlan}) coincide, respectively, 
with the multichannel mesoscopic Josephson equation \cite{Kul75,Bee92} and
the Landauer-B\"{u}ttiker formula \cite{But85};
both formulas are multiplied by a factor of two due to the additional
(valley) degeneracy in graphene.

\subsection{Mode-matching method}
For the Corbino geometry, 
the transmission probabilities in Eqs.\ (\ref{ijoth}) and (\ref{rnlan}),
each of which is attributed to the $j$-th normal mode, with the total
angular momentum quantum number $j=\pm{}1/2,\pm{}3/2,\dots$, 
can be found by solving the 
scattering problem for the Dirac equation, $\mathcal{H}_0\Psi=E\Psi$.
Because of the symmetry of the problem, we can search for the wave function
in the form
\begin{equation}
  \Psi_j(r,\varphi)=e^{i(j-1/2)\varphi}
  \begin{pmatrix} \chi_a \\ \chi_be^{i\varphi} \end{pmatrix}, 
\end{equation}
with the components $\chi_a=\chi_a(r)$, $\chi_b=\chi_b(r)$.
Substituting the above into the Dirac equation immediately brings us to the
system of ordinary differential equations
\begin{align}
  \chi_a' &= \frac{j-1/2}{r}\chi_a
  + i\,\frac{E-V(r)}{\hbar{}v_F}\chi_b,
  \label{phapri} \\
  \chi_b' &= i\,\frac{E-V(r)}{\hbar{}v_F}\chi_a
  - \frac{j+1/2}{r}\chi_b,
  \label{phbpri} 
\end{align}
where primes denote derivatives with respect to $r$.

For the leads, $r<r_1$ or $r>r_2$, the electrostatic potential energy
is constant, $V(r)=-V_0$.
Assuming $E>-V_0$ (electron doping), we can find the solutions
$\chi_j=(\chi_{j,a},\chi_{j,b})^T$ analytically,
\begin{equation}
  \label{chipm}
  \chi_j^{(+)}= 
  \begin{pmatrix}H_{j-1/2}^{(2)}(Kr)\\ iH_{j+1/2}^{(2)}(Kr)\end{pmatrix},
  \ \ \ \ 
  \chi_j^{(-)}= 
  \begin{pmatrix}H_{j-1/2}^{(1)}(Kr)\\ iH_{j+1/2}^{(1)}(Kr)\end{pmatrix}, 
\end{equation} 
where the upper index $(\pm)$ marks the incoming (i.e., propagating
from $r=0$) or the outgoing (propagating from $r=\infty$) wave,  
$H_\nu^{(1)}(\rho)$ [$H_\nu^{(2)}(\rho)$] is the Hankel function of the
first [second] kind, and $K=|E+V_0|/(\hbar{}v_F)$.
Using the above as a~basis set and assuming the scattering from $r=0$,
we write down full wavefunctions (corresponding to a~given $j$) for the
two leads, 
\begin{align}
  \chi_j^{({\rm inner})} &=
  \chi_j^{(+)} + r_j \chi_j^{(-)}, & r<r_1,
  \label{chinner}
  \\
  \chi_j^{({\rm outer})} &= t_j\chi_j^{(+)}, & r>r_2,
  \label{chouter}
\end{align}
where we have introduced the reflection ($r_j$) and
transmission coefficient ($t_j$). 

For the disk area, $r_1<r<r_2$, Eqs.\ (\ref{phapri}), (\ref{phbpri})
typically need to be integrated numerically.
The details of the calculations will be given later; it is now
sufficient impose the boundary conditions
$\left.\chi_j^{(A,B)}\right|_{r-r_1}=(1,\pm{}1)^T$ for the two
linearly-independent solutions, allowing us to write down
\begin{equation}
  \label{chdisk}
  \chi_j^{({\rm disk})}=A_j\chi_j^{(A)}
  + B_j\chi_j^{(B)}, 
\end{equation}
where $A_j$, $B_j$, are arbitrary complex coefficients.

The matching conditions for $r=r_1$ and $r=r_2$ bring us to the linear
system of equations for $A$, $B$, $r_j$, and $t_j$,
\begin{widetext}
\begin{equation}
  \label{lsysABrt}
  \left[
    \begin{matrix}
      \chi_{j,a}^{(-)}(r_1) & -\chi_{j,a}^{(A)}(r_1)
      & -\chi_{j,a}^{(B)}(r_1) & 0 \\
      \chi_{j,b}^{(-)}(r_1) & -\chi_{j,b}^{(A)}(r_1)
      & -\chi_{j,b}^{(B)}(r_1) & 0 \\
      0 &  -\chi_{j,a}^{(A)}(r_2)  & -\chi_{j,a}^{(B)}(r_2)
      & \chi_{j,a}^{(+)}(r_2) \\
      0 &  -\chi_{j,b}^{(A)}(r_2)  & -\chi_{j,b}^{(B)}(r_2)
      & \chi_{j,b}^{(+)}(r_2) \\
    \end{matrix}
  \right]
  \left[
    \begin{matrix}
      r_j \\ A_j \\ B_j \\ t_j \\
    \end{matrix}
  \right]
  =
  \left[
    \begin{matrix}
      -\chi_{j,a}^{(+)}(r_1) \\
      -\chi_{j,b}^{(+)}(r_1) \\
      0 \\
      0 \\
    \end{matrix}
  \right], 
\end{equation}
\end{widetext}
where we have explicitly written the spinor components of relevant
wavefunctions appearing on the right-hand sides of Eqs.\ (\ref{chinner}),
(\ref{chouter}), and (\ref{chdisk}). 

Solving Eq.\ (\ref{lsysABrt}), one finds the transmission amplitude $t_j$
for a~given $j$ and $E=\mu$,  
and the corresponding transmission probability $T_j(\mu)=|t_j|^2$.
The supercurrent $I(\theta)$ and the normal-state resistance $R_N$ are then
determined from Eqs.\ (\ref{ijoth}) and (\ref{rnlan}) by summing over
the modes.

\begin{table*}[tb]
\caption{ \label{icrntable}
  The normal-state conductance $R_N^{-1}$, the product of critical current
  $I_c$ and $R_N$, and the skewness of the current-phase relation $S$
  for the Corbino-Josephson setup in graphene in the high-doping limit
  ($|\mu|\gg{}\hbar{}v_F/r_1$), and at the Dirac point ($\mu=0$), for
  different values of the disk radii ratio $r_2/r_1$. 
  The Sharvin and pseudodiffusive conductance is given by
  $G_{\rm Sharvin}=2g_0j_{\rm max}$, with $g_0=4e^2/h$ and
  $j_{\rm max}=r_1|\mu|/(\hbar{}v_F)$, 
  and $G_{\rm diff}=2g_0/\ln(r_2/r_1)$ (respectively).
  The analytic (or asymptotic) results are given if available. 
}
\begin{tabular}{c|ccc|ccc}
\hline\hline
  $\,r_2/r_1\,$  & $\,R_N^{-1}/G_{\rm Sharvin}\,$ & $\,I_cR_Ne/\Delta_0\,$
    & $\ S\ $ & $\,R_N^{-1}/G_{\rm diff}\,$ & $\,I_cR_Ne/\Delta_0\,$
    & $\ S\ $ \\
      &  \multicolumn{3}{c|}{$|\mu|\gg{}\hbar{}v_F/r_1$}
      &  \multicolumn{3}{c}{$\mu=0$}  \\
\hline
 $1$     &  $\pi/4$     &  $2.42851$  & $\,\ 0.416008\,\ $
   & $1$ & $2.08207$ & $\ 0.254796\ $ \\
 $1.1$   &  $0.806375$  &  $2.43099$  &  $0.415441$
   & $1.00000$ & $2.08207$ & $0.254796$ \\
 $1.25$  &  $0.821914$  &  $2.43436$  &  $0.414544$
   & $1.00000$ & $2.08207$ & $0.254796$ \\
 $2$     &  $0.846108$  &  $2.44141$  &  $0.412430$
   & $0.999963$ & $2.07168$ & $0.240120$ \\
 $5$     &  $0.856558$  &  $2.44504$  &  $0.411246$
   & $0.946965$ & $1.86653$ & $0.122364$ \\
 $10$    &  $0.857949$  &  $2.44554$  &  $0.411080$
   & $0.770470$ & $1.72591$ & $0.063076$ \\
 $\infty$  & $4\!-\!\pi$ & $2.44571$  &  $0.411023$
   & $\,\ \simeq{}4\ln x/x\,^{a)}\,\ $ & $\pi/2$ & $0$ \\
\hline\hline
\multicolumn{7}{l}{
$^{a)}$Asymptotic behavior, $x=r_2/r_1$.
} \\
\end{tabular}
\end{table*}

\section{Rectangular potential barrier}
\label{recbar}

\subsection{Analytic solution}
For a~rectangular barrier of infinite height, corresponding to
$m\rightarrow\infty$ and $V_0\rightarrow\infty$ in Eq.\ (\ref{vrmpot}),
solutions for the leads ($r<r_1$ or $r>r_2$), see Eq.\ (\ref{chipm}), are
replaced with asymptotic forms,
\begin{equation}
  \chi^{(+)}\simeq \frac{e^{iKr}}{\sqrt{r}}
  \begin{pmatrix} 1 \\ 1 \end{pmatrix},
  \ \ \ \ 
  \chi^{(-)}\simeq \frac{e^{-iKr}}{\sqrt{r}}
  \begin{pmatrix} 1 \\ -1 \end{pmatrix}. 
\end{equation}
Solution for the disk ($r_1<r<r_2$) can be written as \cite{Ryc09}
\begin{equation}
  \label{chdiskbzer}
  \chi_j^{({\rm disk})}=A_j
  \begin{pmatrix}H_{j-1/2}^{(2)}(kr)\\ i\eta{}H_{j+1/2}^{(2)}(kr)\end{pmatrix}
  + B_j
  \begin{pmatrix}H_{j-1/2}^{(1)}(kr)\\ i\eta{}H_{j+1/2}^{(1)}(kr)\end{pmatrix}, 
\end{equation}
where $k=|E|/(\hbar{}v_F)$, the doping sign
$\eta=\mbox{sgn}\,E=\pm{}1$ (with $\eta=+1$ indicating electron doping
and $\eta=-1$ indicating hole doping).

In turn, solving Eq.\ (\ref{lsysABrt}) leads to \cite{wronfoo}
\begin{equation}
\label{tjhank}
  T_j=|t_j|^2=
  \frac{16}{\pi^2{}k^2{}r_1{}r_2}\,
  \frac{1}{\left[\mathfrak{D}_{j}^{(+)}\right]^2
    + \left[\mathfrak{D}_{j}^{(-)}\right]^2},
\end{equation}
where
\begin{align}
\label{ddnupm}
  \mathfrak{D}_{j}^{(\pm)} &= \mbox{Im}\left[ 
    H_{j-1/2}^{(1)}(kr_1)H_{j\mp{}1/2}^{(2)}(kr_2)\right. \nonumber\\
    & \ \ \ \ \ \ \ \ \ \ 
    \pm \left.H_{j+1/2}^{(1)}(kr_1)H_{j\pm{}1/2}^{(2)}(kr_2)
    \right]. 
\end{align}
Probably, the most surprising feature of the above result is that
taking the limit of $V_0\rightarrow\infty$ does not give
$T_j\rightarrow{}0$ for all $j$-s; instead, there is a~set of $T_j\sim{}1$ 
for $|j|\lesssim{}kR_{\rm i}$. (The corresponding discussion for the
Corbino disk in 2DEG can be found in Ref.\ \cite{Ryc09}.)  
Also, in the $V_0\rightarrow\infty$ limit, the physical size of the
system is irrelevant (as long as the short-junction condition,
$r_2-r_1\ll{}\xi_0$, is satisfied), and the discussion can be carried out
referring to the dimensionless parameters $r_2/r_1$ and
$\mu{}r_1/(\hbar{}v_F)$. 

\begin{figure*}[t]
\includegraphics[width=\linewidth]{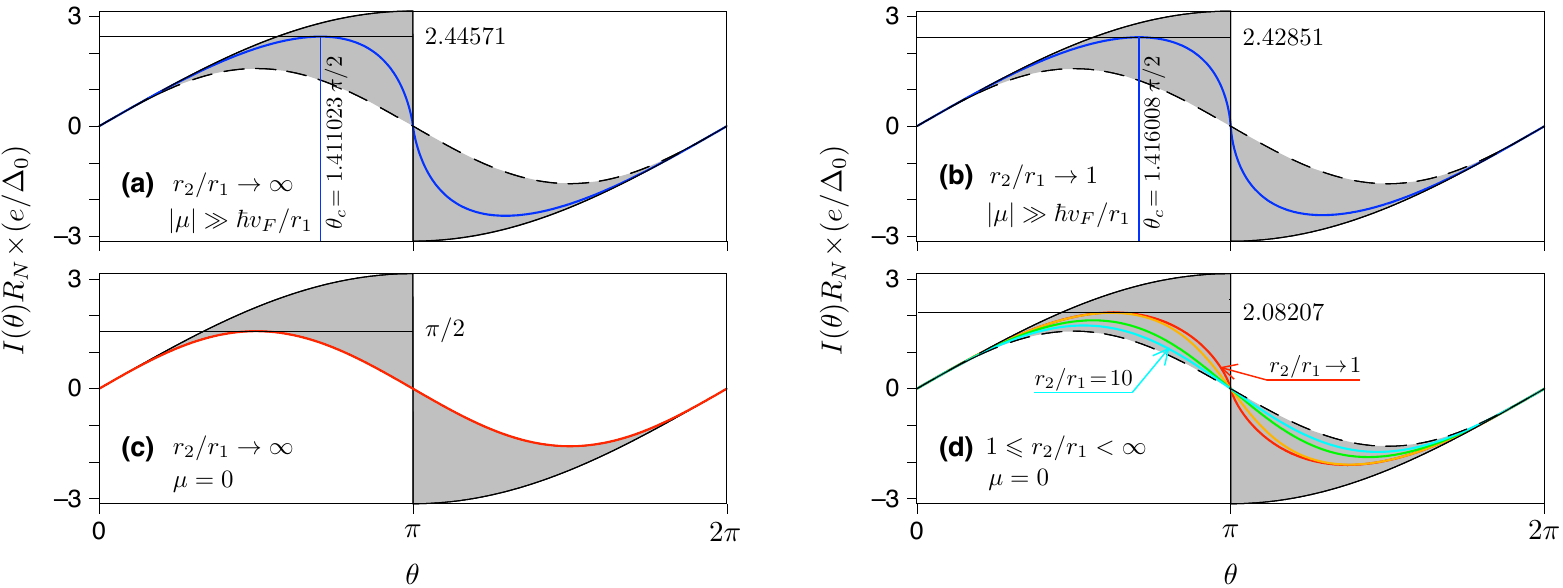}
\caption{ \label{ithdisk12}
  (a)--(d) Current-phase relation for the Corbino-Josephson setup in graphene
  in the case of rectangular potential barrier and infinitely-doped leads,
  corresponding to 
  $m\rightarrow\infty$ and $V_0\rightarrow\infty$ in Eq.\ (\ref{vrmpot}).
  (a,b) The high-doping limit ($|\mu|\gg{}\hbar{}v_F/r_1$),
  (c,d) the Dirac point ($\mu=0$).
  The radii ratio $r_2/r_1$ is specified at each panel.
  Results obtained from Eq.\ (\ref{ithsubsh}) are displayed with
  blue thick lines in (a,b);
  red thick line visualizes Eq.\ (\ref{ijotunn}) in (c) or
  Eq.\ (\ref{ithpdiff}) in (d).
  Remaining color thick lines in (d) visualize Eq.\ (\ref{ijoth}) with
  the probabilities $T_j(0)$ given by Eq.\ (\ref{tjzero}) calculated
  for $r_2/r_1=2$ (orange), $r_2/r_1=5$ (green), and $r_2/r_1=10$ (cyan).
  Thin black lines in all panels visualize the tunneling limit,
  see Eq.\ (\ref{ijotunn})
  (dashed lines), and the ballistic limit, see Eq.\ (\ref{ijoball})
  (solid lines). 
}
\end{figure*}

\subsection{The Dirac point}
In the limit of $k\rightarrow{}0$, Eqs.\ (\ref{tjhank}), (\ref{ddnupm})
simplify to
\begin{equation}
\label{tjzero}
  T_j(0) = \frac{1}{\cosh^2[j\ln(r_2/r_1)]}=\frac{4}{(r_2/r_1)^j+(r_1/r_2)^j}. 
\end{equation}
For $r_2/r_1\rightarrow{}1$ (the narrow-disk limit), one can approximate
the summations in Eqs.\ (\ref{ijoth}) and (\ref{rnlan})
by integrations over continuous $-\infty{}<j<\infty$, reproducing the results
reported in Refs.\ \cite{Tit06,Mog06} for the Dirac point ($\mu=0$)
\begin{equation}
\label{ithpdiff}
  I(\theta)=\frac{e\Delta_0}{\hbar}\frac{4}{\ln(r_2/r_1)}
  \cos(\theta/2)\mbox{artanh}[\sin(\theta/2)],  
\end{equation} 
and
\begin{equation}
\label{rnpdiff}
  R_N^{-1}=\frac{8e^2}{h}\frac{1}{\ln(r_2/r_1)} \equiv{} G_{\rm diff},  
\end{equation}
with the sample aspect ratio $W/L$ replaced by $2\pi/\ln(r_2/r_1)$. 
Finding the maximum of $I(\theta)$ at $\theta=\theta_c$ numerically, we get
\begin{equation}
\label{icrnpdiff}
I_cR_N\frac{e}{\Delta_0}=2.08207 \ \ \ \ 
\text{and }\ \ \ 
S=0.254796, 
\end{equation}
with the skewness of the current-phase relation $S=2\theta_c/\pi-1$.
We further notice that the normal-state conductance ($G_{\rm diff}$) given in
Eq.\ (\ref{rnpdiff}) corresponds to the case of diffusive disk characterized
by the universal conductivity of $\sigma_0=(4/\pi)\,e^2/h$. 

For the wide-disk limit $r_2\gg{}r_1$ and for $\mu=0$, the quantities
given by Eqs.\ (\ref{ijoth}) and (\ref{rnlan}) are governed by two
equivalent modes with $j=\pm{}1/2$, for which $T_j\simeq{}4r_1/r_2\ll{}1$.
Subsequently, Eq.\ (\ref{ijoth}) can be linearized in $T_j$,  with the
result
\begin{equation}
\label{ijotunn}
  I(\theta)\simeq\frac{e\Delta_0}{2\hbar}\sum_{j=\pm{}1/2}
  T_j\sin\theta = \frac{\pi\Delta_0}{2e}R_N^{-1}\sin\theta, 
\end{equation}
giving the characteristics of standard Josephson tunneling (SJT)
\cite{Bee92}, i.e.,  
\begin{equation}
\label{icrntunn}
  I_cR_N\frac{e}{\Delta_0}=\frac{\pi}{2} \ \ \ \ 
\text{and }\ \ \ 
S=0.  
\end{equation}
In effect, the wide-disk (or narrow-opening) geometry, $r_2\gg{}r_1$,
restores SJT
features near the Dirac point, in contrast to the narrow-disk limit
$r_2/r_1\rightarrow{}1$, 
for which graphene-specific multimode Dirac-Josephson tunnelling (MDJT) 
is apparent, see Eqs.\ (\ref{ithpdiff}) and (\ref{rnpdiff}).
These two ranges illustrate different realizations of transport via evanescent
waves in undoped graphene, which --- depending on the sample geometry ---
may either decay exponentially, with multiple almost-equivalent modes
(the $r_2/r_1\rightarrow{}1$ case, or the rectangular sample \cite{Ryc26a}),
or show power-law decay (the $r_2\gg{}r_1$ case) with only two
(fourfold-degenerate) dominant modes. (For a~discussion of transport
characteristics in the case of normal-metal leads, see Ref.\ \cite{Ryc09}). 

\begin{figure}[t]
\includegraphics[width=0.7\linewidth]{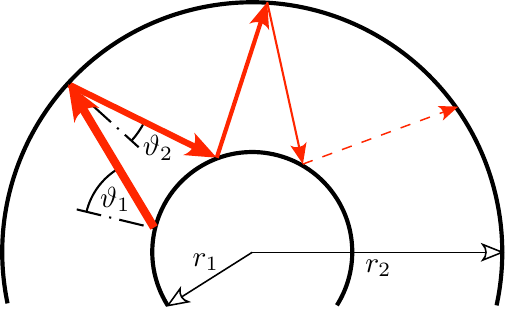}
\caption{ \label{scattering2}
Propagation between scattering on interfaces at $r=r_1$ and $r=r_2$ 
separating the disk area ($r_1<r<r_2$) and the leads ($r<r_1$ or $r>r_2$). 
Angular momentum conservation implies that incident angles for a~particle, 
$\vartheta_1$, $\vartheta_2$, are same for all scatterings on a~given
interface and 
satisfy $|\vartheta_2|<|\vartheta_1|$. 
}
\end{figure}

\subsection{The high-doping limit} 
For the high-doping limit ($|\mu|\gg{}\hbar{}v_F/r_1$), we adopt here 
the approximation technique presented in Ref.\ \cite{Ryc22}.
In brief, it is sufficient to restrict the discussion to
$-j_{\rm max}\leqslant{}j\leqslant{}j_{\rm max}$, with
$j_{\rm max}=r_1|\mu|/(\hbar{}v_F)$, and to replace the sums in
Eqs.\ (\ref{ijoth}) and (\ref{rnlan}) with integrals over $j$.
What is more, due to the presence of two collinear interfaces separating
the sample and the leads, first at $r=r_1$ and second at $r=r_2$, 
the transmission probability $T_j$ can be approximated 
using the double-contact formula \cite{Dat97},
\begin{equation}
\label{ttjdatta}
  T_j\simeq{}T_{j,\phi} = \frac{T_1T_2}{1+R_1{}R_2-2\sqrt{R_1{}R_2}\cos{\phi}}, 
\end{equation}
where the transmission and reflection on the two interfaces read 
\begin{align}
  T_1(j)&=\frac{2\sqrt{1-(j/j_{\rm max})^2}}{1+\sqrt{1-(j/j_{\rm max})^2}}, \\
  T_2(j) &= T_1(jr_1/r_2), \\
  R_l(j)&=1-T_l(j),\ \ \ \  l=1,2,  
\end{align}
and $\phi$ is  a~phase gained between the scattering
events, later assumed to be a~random.
Referring to the picture of classical trajectories, the incident angles, 
$\vartheta_1$ and $\vartheta_2$, at the interfaces at $r=r_1$ and $r=r_2$
(see Fig.\ \ref{scattering2}), 
depend on the quantum number $j$, such that $\sin\vartheta_1=j/j_{\rm  max}$
and $\sin\vartheta_2=(r_1/r_2)\,j/j_{\rm  max}$. 
Subsequently, the phase $\phi=\phi(j)$ is a~function of the distance
traveled between the scattering events.
It can be shown that, in the multimode range ($j_{\rm max}\gg{}1$), when
varying $j$, e.g., by $\delta{}j=1$, the variations of $T_1(j)$ and $T_2(j)$
in Eq.\ (\ref{ttjdatta}) 
are of the order of $\sim{}1/j_{\rm max}$, whereas the variation of $\phi(j)$
is of the order of unity. 
For this reason, physical properties expressed as sums of the form
$\sum_jf(T_j)$, with $T_j\simeq{}T_{j,\phi}$ and $f(x)$ being the analytic
function, can be approximated
by taking both the integral over $j$ and the average over $\phi$. 
Later in this paper, 
such an approximation is confronted with the results following from
the mode-matching method for the DBdG equation, and with the tight-binding
simulations. 
(We further notice that --- in a~real system --- some other factors,
including
the disorder, deformation-induced gauge fields, or fluctuations of $\mu$
during the measurement process, can effectively induce the randomness of
the phase $\phi$.) 

In turn, the Josephson current 
\begin{equation}
\label{ithsubsh}
I(\theta)\simeq{}\frac{e\Delta_0}{\hbar}
\int_{-j_{\rm max}}^{j_{\rm max}}dj\,\frac{1}{2\pi}\int_{-\pi}^{\pi}d\phi\, 
\frac{T_{j,\phi}\sin\theta}{\sqrt{1-T_{j,\phi}\sin^2(\theta/2)}}, 
\end{equation}
where the last integration represents the averaging over $\phi$,  
uniformly-distributed over the range $-\pi\leqslant{}\phi\leqslant{}\pi$. 

For the normal-state resistance, the analogous integrations can be performed
analytically, leading to
\begin{widetext}
\begin{align}
  R_N^{-1} &= \,2g_0\int_0^{j_{\rm max}} dj{}\,\frac{1}{2\pi}\int_{-\pi}^{\pi}d\phi\,
  T_{j,\phi}  =
  G_{\rm Sharvin}\,
  \frac{(2c+\frac{1}{c})\arcsin{}c+3\sqrt{1-c^2}-\frac{\pi}{2}(c^2+2)}{1-c^2},
  \label{ggdiskicoh}
\end{align}
\end{widetext}
where we have defined $g_0=4e^2/h$ (the conductance quantum for graphene),
used parity of Eq.\ (\ref{ttjdatta}) upon
$j\leftrightarrow{}-j$ and $\phi\leftrightarrow{}-\phi$ to shrink the
integration ranges, introduced $G_{\rm Sharvin}=2g_0j_{\rm max}$ being
the Sharvin conductance for a~disk, and defined the inverse radii ratio
$c=r_1/r_2<1$.
As a~closed-form expression for the current-phase relation $I(\theta)$
is unavailable, the results presented below follow from direct numerical
integrations over $dj$ and $d\phi$ in Eq.\ (\ref{ithsubsh}). 
Alternatively, one can represent the right-hand side of Eq.\ (\ref{ithsubsh})
as a~power series of $T_{j,\phi}$ and utilize the analytic expressions for 
charge-transfer cumulants derived in Ref.\ \cite{Ryc25}.
(For more details, see Appendix~\ref{apptns}). 

In particular, for the narrow-disk limit ($r_2/r_1\rightarrow{}1$), the
two interfaces become equivalent ($T_1=T_2$ and $R_1=R_2$), and the
numerical maximization for Eq.\ (\ref{ithsubsh}), with respect to
$\theta$, reproduces the results presented in Ref.\ \cite{Ryc26a}, 
\begin{equation}
\label{icrnsubsh}
I_cR_N\frac{e}{\Delta_0}\simeq{}2.42851 \ \ \ \ 
\text{and }\ \ \ 
S=0.416008, 
\end{equation}
with $R_N^{-1}=(\pi/4)\,G_{\rm Sharvin}$.
The value of the prefactor, $\pi/4<1$, justifies the notion of the 
{\em sub-Sharvin transport regime}. 

In the opposite, wide disk limit ($r_2\gg{}r_1$), transmission via the
second interface becomes perfect ($T_2=1$, $R_2=0$), and the value of
$T_{j,\phi}$ given by Eq.\ (\ref{ttjdatta}) is $\phi$--independent, leading to
the result of 
\begin{equation}
\label{icrnsubshwide}
I_cR_N\frac{e}{\Delta_0}\simeq{}2.44571 \ \ \ \ 
\text{and }\ \ \ 
S=0.411023, 
\end{equation}
with $R_N^{-1}=(4-\pi)\,G_{\rm Sharvin}$.
(Notice that the prefactor, given explicitly in Eq.\ (\ref{ggdiskicoh})
for arbitrary $c=r_1/r_2$, only weakly depends on the geometry.) 

It must be noticed that the dimensionless characteristics given in Eqs.\
(\ref{icrnsubsh}) and (\ref{icrnsubshwide}) are numerically very close to
each other;
together with the so-called {\em pseudodiffusive} values given in
Eq.\ (\ref{icrnpdiff}), they define the borders of graphene-specific
multimode Dirac-Josephson tunneling (MDJT) on the $(I_cR_N,S)$ diagram
discussed later in this paper.
Although the pseudodiffusive and sub-Sharvin transport regimes are
characterized by different distributions of transmission eigenvalues
\cite{Ryc25}, physical properties probed in the S-g-S setup are relatively
close to each other, yet noticeably diferent from those occurring for SJT
and ballistic regimes (see below); at the same time, behavior in the whole
MDJT range is directly linked to the peculiar properties following from the
conical dispersion relation, either the transmission via evanescent waves
(at the Dirac point) or reduced transmission on sample-lead interface
(at the high-doping limit). 

In Table \ref{icrntable}, we list the values of $I_cR_N$ and $S$ for
six finite values of $r_2/r_1=1.1\div{}10$, both for the high-doping
limit ($|\mu|\gg{}\hbar{}v_F/r_1$) and for the Dirac point ($\mu=0$),
together with the limiting results for $r_2/r_1\rightarrow{}1$ and
$r_2/r_1\rightarrow{}\infty$ presented above. 

The corresponding current-phase relations $I(\theta)$, obtained numerically
from Eqs.\ (\ref{ithpdiff}), (\ref{ijotunn}) and (\ref{ithsubsh}), are
displayed in Figs.\ \ref{ithdisk12}(a)--(c). 
Remarkably, the results for the Dirac point ($\mu=0$) and finite $r_2/r_1$,
see Fig.\ \ref{ithdisk12}(d) [solid lines], obtained by the numerical
summation over $j$-s in Eqs.\ (\ref{ijoth}) and (\ref{rnlan}) with $T_j(0)$
given by Eq.\ (\ref{tjzero}), demonstrate gradual evolution --- with
increasing $r_2/r_1$ --- between the limiting curves for $r_2/r_1\rightarrow{}1$
(red solid line) and for the SJT limit, $r_2/r_1\rightarrow{}\infty$ [black
dashed line]. (Notice that only the lines for $r_2/r_1=2$, $5$, and $10$
are displayed, since for $r_2/r_1\leqslant{}1.25$ the overlap with the
$r_2/r_1\rightarrow{}1$ results are almost perfect.)

Also in Figs.\ \ref{ithdisk12}(a)--(d), we display [with thin solid lines]
the current-phase relation for a perfect ballistic system (or Sharvin
contact). For such a~system, transmission eigenvalues are equal to either $0$
or $1$, and can be ordered such that
\begin{equation}
  T_n=\begin{cases} 1, & 0\leqslant{}n<N_0,\\
  0, & N_0\leqslant{}n<N, 
\end{cases}
\end{equation}
with the number of open channels $N_0\ll{}N$.
Substituting the above to Eqs.\ (\ref{ijoth}) and (\ref{rnlan}) brings us
to 
\begin{equation}
\label{ijoball}
  I(\theta)=N_0\frac{e\Delta_0}{\hbar}
  \sin(\theta/2)\,{\rm sgn}(\cos\theta/2), 
\end{equation}
where ${\rm sgn}(x)$ is the sign function, and
\begin{equation}
  R_N^{-1}=N_0\frac{4e^2}{h}. 
\end{equation}
In effect, for the ballistic Josephson effect (BJE) 
\begin{equation}
\label{icrnball}
  I_cR_N\frac{e}{\Delta_0}=\pi \ \ \ \ 
\text{and }\ \ \ 
S=1.  
\end{equation}

Eqs.\ (\ref{ijoball}) and (\ref{icrnball}) refer to idealized situation
with fully-closed or fully-open channels only.
In a~physical system, intermediate values appear, leading the position
in $(I_cR_N,S)$ coordinates close to, but not perfectly matching, the point
defined in Eq.\ (\ref{icrnball}).
The further analysis of such a~crossover to the BJE range is presented next. 

\begin{figure}[t]
\includegraphics[width=\linewidth]{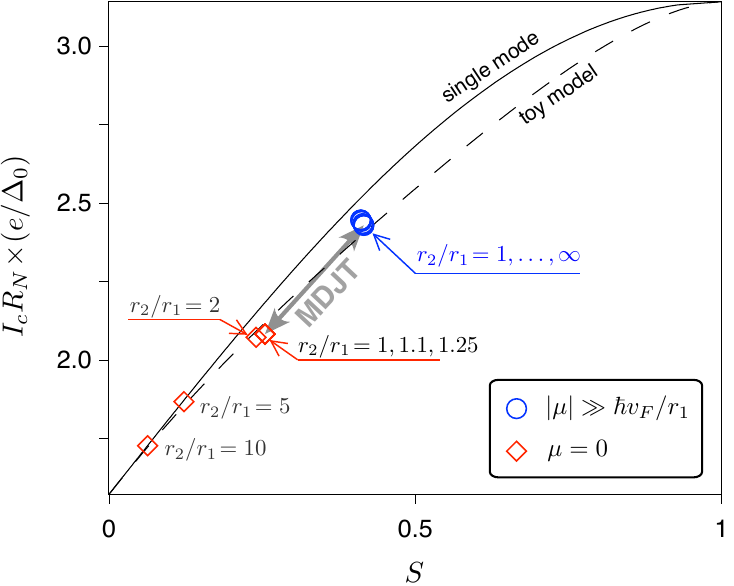}
\caption{ \label{diagIcSdisk}
  Product of critical current and normal-state resistance $I_cR_N$ displayed
  versus skewness of the current-phase relation (datapoints).
  Two datasets represent the results of Table~\ref{icrntable}, for the
  high-doping limit ($|\mu|\gg{}\hbar{}v_F/r_1$) [circles],
  and for the Dirac point ($\mu=0$) [diamonds].
  Double arrow indicates the multimode Dirac-Josephson tunneling (MDJT)
  regime, bounded by the $(I_cR_N,S)$ values given in Eqs.\ (\ref{icrnpdiff})
  and (\ref{icrnsubsh}).
  Black solid line presents the results following from the maximization
  of Eq.\ (\ref{ithsinmod}) for a~single nonzero eigenvalue $0<T\leqslant{}1$; 
  black dashed line depicts the results obtained within the multimode toy
  model defined by Eq.\ (\ref{tttoy}). 
}
\end{figure}

\subsection{The critical current---skewness diagram}
In Fig.\ \ref{diagIcSdisk}, we present the product $I_cR_N$ as a~function
of skewness $S$, using the data of Table \ref{icrntable}.
Hereinafter, the pairs of $(I_cR_N,S)$ defined by Eqs.\ (\ref{icrnpdiff})
and (\ref{icrnsubsh}), are chosen as bounds the MDJT range 
(indicated with a~double arrow).

In order to rationalize our numerical results for the Corbino-Josephson
setup in graphene, we consider (as a~first toy-model)
the case of a~single nonzero eigenvalue,
yielding the current-phase
relation
\begin{equation}
\label{ithsinmod}
I(\theta)=\frac{e\Delta_0}{\hbar}
\frac{T\sin\theta}{\sqrt{1-T\sin^2(\theta/2)}}, 
\ \ \ 
\text{with }
\ \ \ 
0<T\leqslant{}1, 
\end{equation}
and $R_N^{-1}=(4e^2/h)\,T$; see Eqs.\ (\ref{ijoth}) and (\ref{rnlan}).
Straightforward maximization of $I(\theta)$ with respect to $\theta$
leads to the dependence of $I_cR_N$ versus $S$ presented with a~black solid
line in Fig.\ \ref{diagIcSdisk}. 
The characteristics of such a~{\em single-mode Josephson
junction} are quite distant from the results for the Corbino-Josephson
setup, except from the data for $\mu=0$ and $r_2/r_1\gtrsim{}5$,
where the system enters the SJT range. 

To construct an alternative {\em multimode toy model} that parametrizes
the tunneling-to-ballistic crossover, we propose the following
transmission-angular momentum dependence
\begin{equation}
\label{tttoy}
  T_{j}^{(\Theta)}=
  \frac{1}{e^{j-\Theta}+1} - \frac{1}{e^{j+\Theta}+1}
  \ \ \ \ \ \ (\Theta>0).    
\end{equation}
It is also supposed that $j_{\rm max}\gg{}1$, and thus the summations in Eqs.\
(\ref{ijoth}) and (\ref{rnlan}) are replaced by integrations over
$-\infty<j<\infty$. 
For instance, $\Theta\rightarrow{}0$ reproduces the tunneling limit, with
the values of $I_cR_N$ and $S$ given by Eq.\ (\ref{icrntunn}), whereas
$\Theta\gg{}1$ corresponds to the ballistic limit. 
(To be more specific, the value of $I_cR_N$ given in Eq.\ (\ref{icrnball}) is
reproduced with an accuracy better than $1\%$ for $\Theta\geqslant{}1200$; 
the same applies for $S$, starting from $\Theta=4\cdot{}10^4$.)  

The functional dependence of $I_cR_N$ on $S$, which follows from 
Eq.\ (\ref{tttoy}), is also visualized in Fig.\ \ref{diagIcSdisk} 
(dashed line). 
We find that such dependence can be approximated by
\begin{equation}
\label{tttoyicrn}
  I_cR_N\frac{e}{\Delta_0} \simeq
  \frac{\pi}{2}(S+1) + 0.59\,S^{0.89}\left(1-S\right)^{0.70}. 
\end{equation}
The corresponding curve is omitted since it matches the dotted line in Fig.\
\ref{diagIcSdisk} perfectly.
It is worth to pointing here that the model defined by Eq.\ (\ref{tttoy}),
constituting a~phenomenological reasoning partly inspired by the results
for Schr\"{o}dinger electrons tunneling through a~parabolic potential obtained
by Kemble in 1935 \cite{Kem35}, 
is not specific for graphene, so it may also apply to other multimode systems
showing the tunneling-to-ballistic crossover, such as disordered wires
\cite{Mis01}. 

As the current-phase relation for $r_2/r_1=5$, see Fig.\ \ref{ithdisk12}(d)
[green solid line], as well as the corresponding  $(I_cR_N,S)$ points 
in Fig.\ \ref{diagIcSdisk}, 
lay relatively far from both the $r_2/r_1\rightarrow{}1$
and $r_2/r_1\rightarrow{}\infty$ limits, this value is selected for the
numerical analysis of smooth potentials presented in next Section.

\begin{figure}[t]
\includegraphics[width=\linewidth]{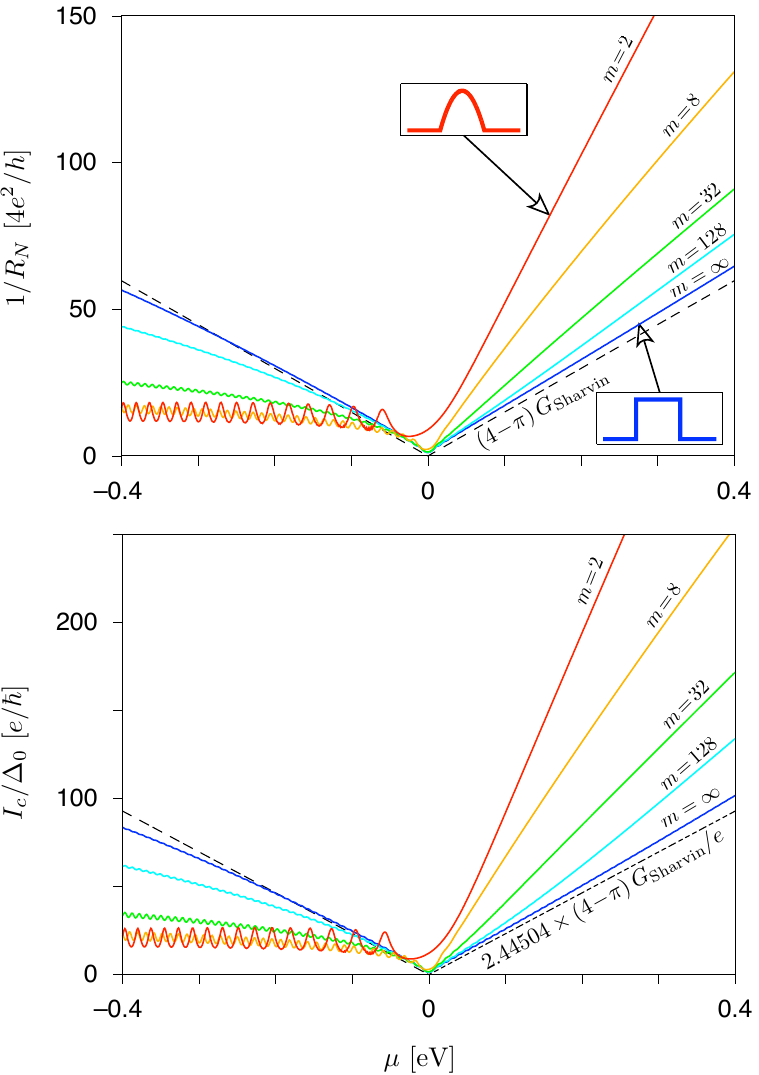}
\caption{ \label{gicR50L200vsEab}
  Normal-state conductance $1/R_N$ (top) and critical current $I_c$ (bottom)
  for the system of Fig.\ \ref{diskmpot} as functions of the chemical
  potential ($\mu=E$).  The parameters are: $r_2=5\,r_1=250\,$nm,
  $V_0=t_0/2=1.35\,$eV. The exponent $m$ in Eq.\ (\ref{vrmpot}) is specified
  for each dataset (solid lines). Insets (top) depict the potential profiles
  for $m=2$ and $m=\infty$. Dashed line depicts the sub-Sharvin 
  conductance and the corresponding critical current for 
  $r_2/r_1\rightarrow\infty$, see Eq.\ (\ref{icrnsubshwide}). 
}
\end{figure}

\begin{figure*}[t]
\includegraphics[width=\linewidth]{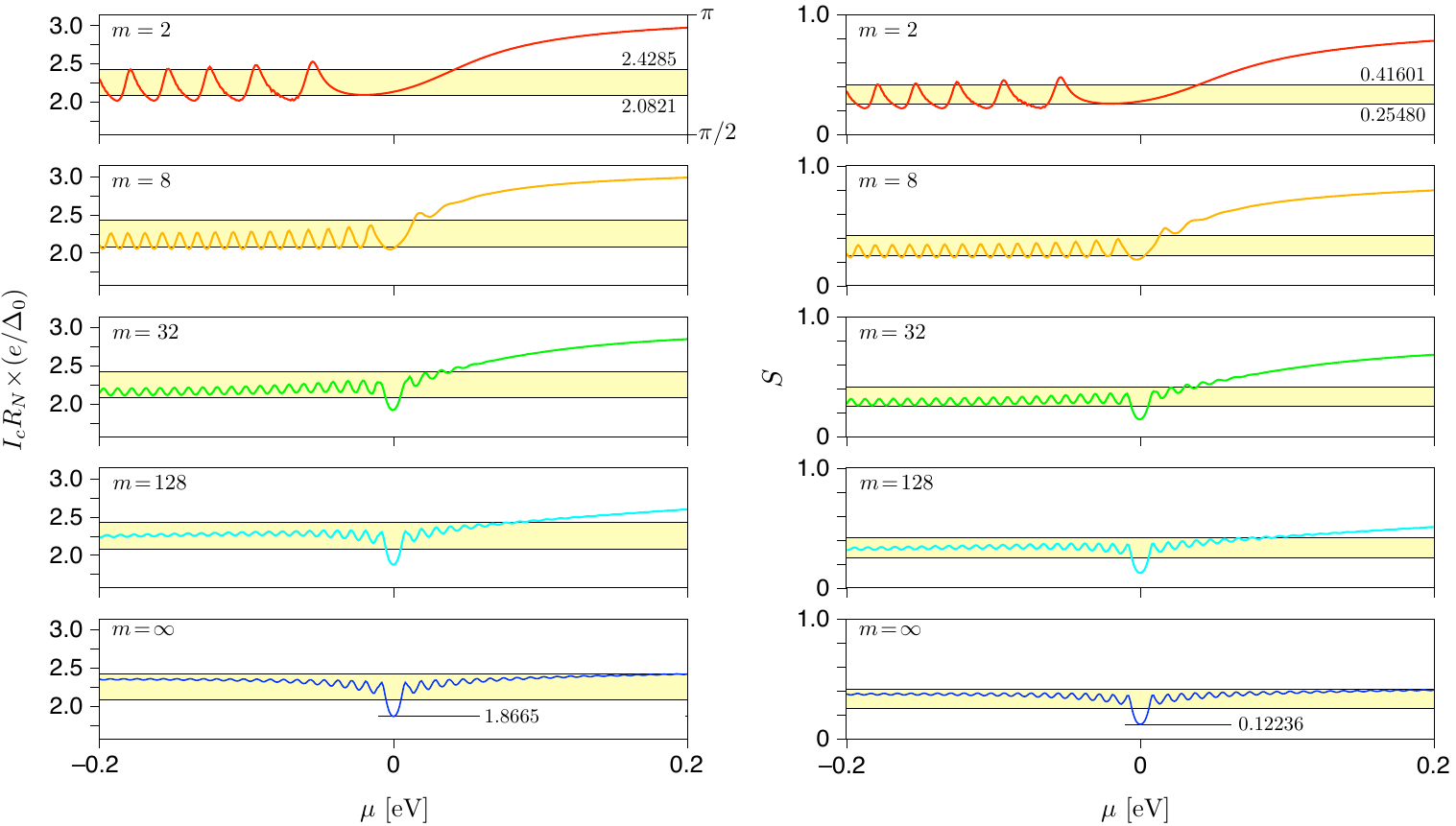}
\caption{ \label{icrnsmax10pan}
  Left: Product $I_cR_N$ for the data shown in Fig.\ \ref{gicR50L200vsEab}.  
  Right: Skewness of the current-phase relation $S$ as a~function of the
  chemical potential for the same system parameters. 
  The exponent $m$ in Eq.\ (\ref{vrmpot}) is varied between the rows. 
  Horizontal lines bordering the yellow areas mark the MDJT range, defined by
  the values for rectangular barrier of an infinite height
  ($m\rightarrow\infty$, $V_0\rightarrow\infty$), corresponding to $\mu=0$ and
  $|\mu|\gg{}\hbar{}v_F/r_1$ for the narrow-disk limit, 
  $r_2/r_1\rightarrow{}1$, see Eqs.\ (\ref{icrnpdiff}) and (\ref{icrnsubsh});
  the values for $r_2/r_1=5$ are also marked in the bottom rows. 
}
\end{figure*}

\begin{figure}[t]
\includegraphics[width=\linewidth]{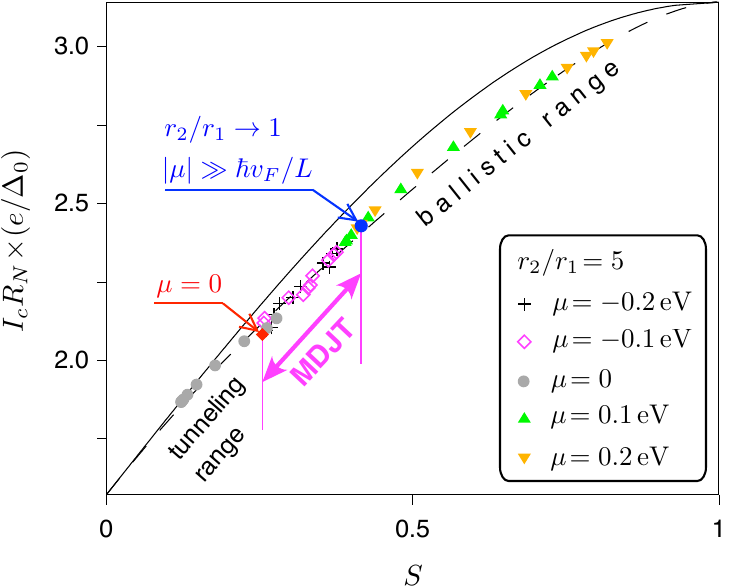}
\caption{ \label{diagIcSmpots}
  Product $I_cR_N$ displayed versus the skewness for smooth potentials
  (datapoints).
  Each dataset contains ten datapoints corresponding to $m=2,4,\dots,512$,
  and $m=\infty$, for one of five fixed values of the chemical potential,
  $\mu=-0.2,\,-0.1,\,\dots,\,0.2\,$eV (see the legend).
  Remaining system parameters are same as in Fig.\ \ref{gicR50L200vsEab}.
  Two additional symbols (full circle and diamond) mark the values given
  in Eqs.\ (\ref{icrnpdiff}) and (\ref{icrnsubsh}), separating
  the crossovers to SJT (lower values), the MDJT regime (middle), 
  and the crossover to BJE (higher values). 
  Solid/dashed black lines are same as in Fig.\ \ref{diagIcSdisk}. 
}
\end{figure}

\section{Smooth potentials}
\label{smopot}

\subsection{The system and computation details}
So far, we have focused on the limits of $|\mu|\gg{}\hbar{}v_F/r_1$ and
$\mu\ll{}\hbar{}v_F/r_1$, for which several analytic results are available. 
From now on, the discussion is carried out in physical units,
by looking 
at the finite range of $|\mu|\leqslant{}0.4\,$eV, comparable
with the range accessible for graphene-on-hBN devices \cite{Zen19}. 
As the essential transport characteristics for Corbino disks in graphene
saturate (i.e., become very close to their limiting values for
$r_2/r_1\rightarrow{}1$) for $r_2/r_1\lesssim{}2$ \cite{Ryc10,Ryc22},
we take a~wider disk, with the radii at $r_2=5r_1=250\,$nm,
allowing one to expect the intermediate features between the SJT and MDJT
regimes (see Sec.\ \ref{recbar}).
The inner lead diameter ($2r_1=100\,$nm) exceeds the diameter of nanoscopic
suspended leads fabricated of MoRe \cite{Azi14}. 
Such choice also defines the energy scale that 
separates the weak- and the high-doping regimes at $\hbar{}v_F/r_1=11.5\,$meV. 
The step height in Eq.\ (\ref{vrmpot}) is $V_0=t_0/2=1.35\,$eV, which yields
(for instance) the number of propagating modes in the inner lead as
$216$ for $\mu=-0.1\,$eV and $252$ for $\mu=0.1\,$eV.
The sample length, $r_2-r_1\equiv{}L=200\,$nm, corresponds to
$L/\xi_0=\Delta_0/\varepsilon{}_c\approx{}0.36$ (with the Thouless energy
$\varepsilon_c=\hbar{}v_F/L$) for MoRe leads. 
In turn, the deviations from the short-junction limit, see Eq.\ (\ref{ijoth}),
are expected to be negligible \cite{Dub01}. 

The numerical integration of Eqs.\ (\ref{phapri}) and (\ref{phbpri}) was 
performed utilizing a~standard fourth-order Runge-Kutta (RK4) algorithm, with
a~spatial step of $\Delta{}x=L/80000=2.5\,$pm. 
(For each value of $j$, the wavefunctions analogous to those given in Eqs.\
(\ref{chinner}) and (\ref{chouter}), but
describing scattering from $r=\infty$ towards $r=0$, were used to construct
the linear system analogous to Eq.\ (\ref{lsysABrt}), which 
was then solved 
to find the amplitudes $t_j'$ and $r_j'$, and to check the unitarity of the
scattering matrix $\mathcal{S}$; the above parameters result in an unitarity
error $\epsilon_n=||\mathcal{S}\mathcal{S}^\dagger-I||\lesssim{}10^{-5}$,
with $||M||$
denoting the maximum absolute value of a~matrix element $M_{ij}$ and the
identity matrix $I$, for all cases considered.)  
Summation over the modes in Eqs.\ (\ref{ijoth}), (\ref{rnlan}) was terminated
if $T_n<10^{-6}$. 

\subsection{Transport characteristics}
The evolution of the conductance spectrum with exponent $m$ in Eq.\
(\ref{vrmpot}), which was discussed earlier in Ref.\ \cite{Ryc22}, 
but only for $r_2/r_1\leqslant{}2$, 
is visualized in the top panel of Fig.\ \ref{gicR50L200vsEab}. 
The bottom panel of Fig.\ \ref{gicR50L200vsEab} presents a~similar evolution
of the critical current. 
Remarkably, the behavior of the two physical properties is very similar,  
particularly in the tripolar regime ($\mu<0$). 
For the unipolar regime ($\mu>0$), $I_c$ decays slightly slower than $1/R_N$
with the increasing $m$, particularly for smaller $m$-s.
(Notice that the particle-hole asymmetry is amplified with lowering $m$, in
agreement with the earlier results briefly overviewed in
Sec.\ \ref{modmet}A.) 

A~deeper insight into the system behavior is provided by the evolution
of the product $I_cR_N$ and the skewness $S$, as shown in
Fig.\ \ref{icrnsmax10pan}. 
It is noticeable that the numerical results for smooth potentials and finite
$V_0$ stay in the graphene-specific MDJT range, which is 
defined by the values obtained for $r_2/r_1\rightarrow{}1$
for the rectangular barrier of an infinite height for $\mu=0$ and 
$|\mu|\gg{}\hbar{}v_F/r_1$, see Eqs.\ (\ref{icrnpdiff}) and (\ref{icrnsubsh}),
provided that the system is in the tripolar regime ($\mu<0$).
Even for the lowest considered $m=2$, for which the resonances with
quasi-bound states are well-pronounced \cite{Sil07}, very few data lay
outside the borders of the MDJT range. (Similar behavior occurs for the
rectangular sample, see Ref.\ \cite{Ryc26a}.) 
In contrast, in the unipolar regime ($\mu>0$), $I_cR_N$ evolves
--- with the increasing $m$ --- from
the values close to the ballistic limit, see Eq.\ (\ref{icrnball}),
towards graphene-specific values.

In the vicinity of the Dirac point, $|\mu|\ll{}\hbar{}v_F/r_1$, 
the tunneling limit, see Eq.\ (\ref{icrntunn}),
is approached for high $m$-s; the graphene-specific range is
re-entered for lower $m$-s.
Such a~feature is virtually invisible for both $1/R_N$ and $I_c$, 
see Fig.\ \ref{gicR50L200vsEab} suggesting that intensive quantities,
such as the product $I_cR_N$ and the skewness $S$, are better probes of
graphene-specific features in superconductor-graphene-superconductor systems.
We also notice here that the existing experimental measurements of $S$
for rectangular Josephson junctions in graphene seem to be weakly affected
by contact resistances, for instance Refs.\  
\cite{Eng16,Nan17} report values of $S\approx{}0.2\div{}0.25$ near the Dirac
point, as well as in the tripolar regime. 
Similar results were recently reported for other systems showing the conical
dispersion relation \cite{Sur23}. 

For a~somewhat more detailed view of the data, we present in Fig.\
\ref{diagIcSmpots} the product $I_cR_N$ as a~function of skewness $S$
for five selected values of $\mu=0$, $\pm{}0.1\,$eV, $\pm{}0.2\,$eV, and
ten different values of $m$ ($m=2,4,\dots,512$, and $\infty$) for each $\mu$. 
The two $(I_cR_N,S)$ pairs, defined by Eqs.\ (\ref{icrnpdiff}),
(\ref{icrnsubsh}), and bounding the MDJT regime, are also marked 
(and indicated with arrows). 
Additionally, the single-mode approximation,
see Eq.\ (\ref{ithsinmod}), 
and the results obtained within
the multimode toy model, see Eqs.\ (\ref{tttoy}) and (\ref{tttoyicrn}),  
are visualized in Fig.\ \ref{diagIcSmpots} with solid and dashed lines 
(respectively).

The values of $(I_cR_N,S)$ for smooth potentials are generally quite distant
from the characteristics of a~single-mode Josephson junction (see Fig.\
\ref{diagIcSmpots}); instead, they are much closer 
to the results following from the multimode toy model.
This feature is particularly striking within (and close to) the MDJT regime
(in particular, the values given in Eqs.\ (\ref{icrnpdiff}) and
(\ref{icrnsubsh}) are well-reproduced for $\Theta\approx{}3.4$ and
$\Theta\approx{}6.8$), but also in the BJE limit.
As a~secondary difference between the $\mu>0$ and $\mu<0$ results in the MDJT
regime, we observe that for $\mu=-0.1\,$eV and $-0.2\,$eV the datapoints are
slightly more dispersed around the line corresponding to the multimode model
then for $\mu=0.1\,$eV and $0.2\,$eV. 
This is another manifestation of resonances with quasi-bound states occurring
for lower $m$-s in the tripolar n-p-n doping case. 

For $\mu=0$ and large $m$, where SJT is restored, 
the single-mode model provides a~reasonable approximation for the numerical
results.

\begin{figure*}[t]
\includegraphics[width=\linewidth]{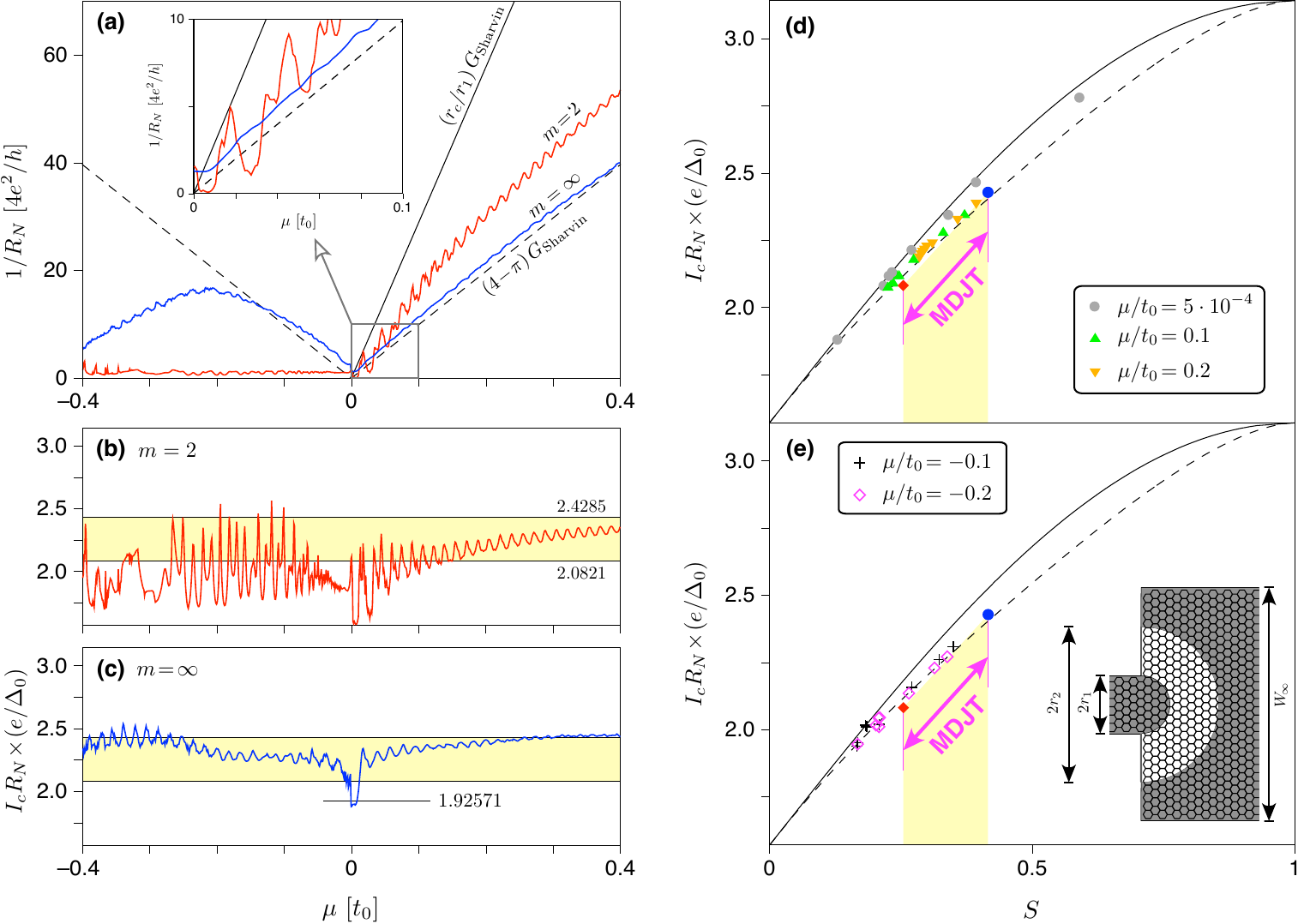}
\caption{ \label{gicdiag700x240}
  (a)--(e) Normal-state conductance $1/R_N$ in the units of $g_0=4e^2/h$ (a),
  the product $I_cR_N$ in the units of $\Delta_0/e$ (b,c)
  as functions of the chemical potential (specified in the units of
  $t_0=2.7\,$eV) and the critical current---skewness diagram (d,e) for
  the half-Corbino disk, shown schematically in (e).
  The system parameters are $W_{\infty}=700\,a_0$, $r_2=4r_1=200\,a_0$.
  (a--c) Color thick lines represent the numerical results the potential
  profiles given by Eq.\ (\ref{vrmpot}) with $m=2$ [red] and $m=\infty$
  [blue]. 
  Black thin lines in (a) depict the ballistic approximation for a~continuous
  system, $(r_c/r_1)\,G_{\rm Sharvin}$ with $G_{\rm Sharvin}=g_0r_1|\mu|/(\hbar{}v_F)$
  [solid] and the sub-Sharvin conductance $(4-\pi)\,G_{\rm Sharvin}$ [dashed].
  (Inset is a~zoom-in, for $0\leqslant{}\mu/t_0\leqslant{}0.1$, of the data
  shown in the main panel.) 
  Horizontal lines bounding the yellow areas (b,c) are same as
  in Fig.\ \ref{icrnsmax10pan}.
  Additional horizontal line in (c) marks the value of $I_cR_N$ 
  for $r_2/r_1=4$ and $V_0,m\rightarrow{}\infty$ in the continuous model,
  see Eq.\ (\ref{icrn0half}). 
  (d,e) Each dataset contains eight datapoints corresponding to
  $m=2,4,\dots,128$, and $m=\infty$, for one of fixed values of the chemical
  potential, i.e., $\mu/t_0=5\cdot{}10^{-4}$, $0.1$, $0.2$ (d), or
  $\mu/t_0=-0.1$, $\mu/t_0=-0.2$ (e) [see the legend]. 
  Solid/dashed black lines are same as in Fig.\ \ref{diagIcSdisk}; 
  two additional symbols (full circle and diamond) bounding the MDJT range 
  are same as in Fig.\ \ref{diagIcSmpots}. 
}
\end{figure*}

\section{Tight-binding simulation}
\label{tbasim}

For the sake of completeness, we present now the results of tight-binding
simulations for the half-Corbino disk --- the system earlier studied in
Ref.\ \cite{Ryc25}. 
Going beyond the Dirac equation, the tight-binding model grasps several
features of more accurate models, such as the trigonal warping of the
dispersion relation, the presence of the van Hove singularity in the density
of states, and the short-wavelength cutoff due to the lattice discretization,
manifesting itself via limited number of modes in the leads. 

\subsection{The Hamiltonian}
Although the Hamiltonians with more distant hopping elements were recently
discussed \cite{Vid22}, here we will limit ourselves to the familiar
single-hopping tight binding Hamiltonian given by 
\begin{equation}
\label{hamtba}
  H = \sum_{i,j,s}t_{ij}c_{i,s}^{\dagger}c_{j,s}+\sum_{i,s}V_in_{is}, 
\end{equation}
where the indices $i$, $j$ run over sites in the honeycomb lattice
of carbon atoms, and $s=\uparrow,\downarrow$ is the spin up/down orientation. 
The hopping-matrix elements $t_{ij}=-t_0$ (with $t_0=2.7\,$eV) if $i,j$ are
nearest neighbors; otherwise, $t_{ij}=0$. 
For the system depicted in Fig.\ \ref{gicdiag700x240}(d), the electrostatic
potential energy $V_j=V({\bf r}_j)$ varies according to Eq.\ (\ref{vrmpot})
in the sample area, and is equal to $-V_0$, with $V_0=t_0/2=1.35\,$eV, 
in the leads.
Since the Hamiltonian given by Eq.\ (\ref{hamtba}) includes nearest-neighbor
hopping elements only, a~small deviation of the potential energy, 
$\delta{}V_j=0.01\,t_0$, is added for outermost edge atoms in the sample area
in order to ensure the physical character of the current density distribution
for the case of $\mu\approx{}0$ \cite{Wim09}. 

Due to the complexity of tight-binding calculations (notice that the
cylindrical symmetry and angular-momentum conservation no longer apply, 
and scattering cannot be described independently for each normal mode, 
as in previous sections; instead, the mode mixing occurs \cite{Ryc25})
the physical size of the system is reduced. 
Namely, we took $W_{\infty}=700\,a_0$ (the lead width) and the radii
$r_1=50\,a_0$, $r_2=200\,a_0$. 
In effect, the energy separating weak- and the high-doping regimes
is $\hbar{}v_F/r_1=46.8\,$meV, 
the number of sites in the sample area
equals $136,035$ (comparing to the total no.\ of $336,000$ sites placed
between the semi-infinite leads at a~distance of $120\sqrt{3}\,a_0\simeq{}
51.1\,$nm).
The details of the computational technique can be found in Ref.\
\cite{Ryc25}.
Similarly as before, the eigenvalues ($T_n$) 
of the matrix ${\boldsymbol t}{\boldsymbol t}^{\dagger}$, where the transmission
matrix ${\boldsymbol t}=(t_{mn})$, $m=1,\dots,N_R$, $n=1,\dots,N_L$, and 
$N_R$ ($N_L$) denotes the number of modes in the right (left) lead, 
are determined for a~given Fermi energy $E$, the normal-state conductance as
well as the Josephson current can be evaluated via Eqs.\ (\ref{ijoth}) and
(\ref{rnlan}).

\subsection{Relation to the continuous model}
It is a~notable feature of the half-disk geometry that considerations for
the continuous model (i.e., starting from the Dirac equation)
\cite{Ryc09,Ryc22}, with the mass confinement \cite{Ber87} and
$V_0,m\rightarrow{}\infty$ in Eq.\ (\ref{vrmpot}), lead to identical
transmission eigenvalues ($T_j$) as given by Eq.\ (\ref{tjhank}), with
the $\mu\rightarrow{}0$ limits $T_j(0)$ given by Eq.\ (\ref{tjzero}), but
the quantum number $j$ takes the positive values,
$j=\frac{1}{2},\frac{3}{2},\dots$.
In effect, the pseudodiffusive values of $I(\theta)$ and $R_N^{-1}$ given by
Eqs.\ (\ref{ithpdiff}) and (\ref{rnpdiff}), need to be divided by a~factor of
$2$; the same applies for the sub-Sharvin results given by Eqs.\
(\ref{ithsubsh}) and (\ref{ggdiskicoh}).
In particular, for $|\mu|\gg{}\hbar{}v_F/r_1$ and $r_2\gg{}r_1$, we get
\begin{equation}
  R_N^{-1}\simeq{}(4-\pi)\,G_{\rm Sharvin},
\end{equation}
with $G_{\rm Sharvin}=g_0r_1|\mu|/(\hbar{}v_F)$ this time. 

We further notice that intensive quantities, such as the product $I_cR_N$
and the skewness $S$ are unaffected upon the disk halving, as long as the
Dirac-equation limit is considered. 
For the Dirac point $|\mu|\ll{}\hbar{}v_F/r_1$ and a~finite $r_2/r_1=4$, 
substituting $T_j(0)$-s given by Eq.\ (\ref{tjzero}) into Eqs.\ (\ref{ijoth}),
(\ref{rnlan}), and summing over $j=\frac{1}{2},\frac{3}{2},...$ leads to
\begin{equation}
\label{icrn0half}
  I_c{}R_N\frac{e}{\Delta_0}= 1.92571, \ \ \ \ 
  S= 0.149323.   
\end{equation}
Similarly as for $r_2/r_1=5$ (see Table~\ref{icrntable}), the above values
are significantly lower than their pseudodiffusive counterparts given in 
Eq.\ (\ref{icrnpdiff}).

Apart from the above-mentioned coinciding predictions for the full disk
and the half disk obtained within the continuous model, the limited
size and the discretness of the tight-binding system introduces several
specific factors (see below); in turn, the simulation should rather be
considered as a~qualitative robustness check of the key results rather than
a~quantitative test of the DBdG equation-based theory. 

\subsection{Numerical results}
The results for our tight-binding simulations are presented in
Fig.\ \ref{gicdiag700x240}.
Since the evolution with the exponent $m$, see Eq.\ (\ref{vrmpot}), is
systematic, we limit the presentation in (a)--(c) to the parabolic ($m=2$)
and rectangular ($m=\infty$) profiles.
In particular, the conductance spectrum in (a) is approximately symmetric
(upon $\mu\leftrightarrow{}-\mu$) for $m=\infty$ if $|\mu|/t_0\lesssim{}0.2$.
For $\mu/t_0\lesssim{}-0.2$, the number of propagating modes in the left
(narrow) lead, $N_L\simeq{}r_1|V_0+\mu|/(\hbar{}v_F)\lesssim{}17$, and the
conductance reduction becomes significant.
Additionally, for $\mu/t_0\gtrsim{}-0.2$, numerical results are close to
the sub-Sharvin values given by $(4-\pi)\,G_{\rm Sharvin}$ [dashed line].

Also in Fig.\ \ref{gicdiag700x240}(a) but for $m=2$, we identify strong
conductance suppression for the tripolar doping ($\mu<0$), where
the resonances with quasi-bound states, earlier observed for the
continuous model (see Fig.\ \ref{gicR50L200vsEab}), are virtually absent. 
This can be attributed partly to the lattice discretization, and partly
to the group velocity mismatch due to trigonal warping.
In contrast, for the unipolar doping ($\mu>0$), the reduction is only 
moderate in comparison to the expected ballistic value \cite{Ryc22}
\begin{equation}
  G_{\rm Sharvin}^{(2)}\approx{}\left(r_c/r_1\right)\,G_{\rm Sharvin}^{(\infty)}, 
\end{equation}
with $G_{\rm Sharvin}^{(\infty)}\equiv{}G_{\rm Sharvin}$,
shown with black solid line.
The effect is less apparent for longer Fermi wavelengths (see the inset),
suggesting that the behavior of the continuous model will presumably be
restored for a~sufficiently large lattice system. 

Remarkably, for the product $I_cR_N$, see Figs.\ \ref{gicdiag700x240}(b,c),
is rather weakly affected by the above-mentioned lattice-related factors.
For $m=\infty$, most of the data --- in the whole range of
$|\mu|/t_0\leqslant{}0.2$ considered --- lay within the MDJT range (yellow
area) defined by the values given in Eqs.\ (\ref{icrnpdiff}) and
(\ref{icrnsubsh}), with the exception of a~small vicinity of $\mu=0$, where
the value given in Eq.\ (\ref{icrn0half}) is closely approached.
For $m=2$, the oscillations of Fabry-P\'{e}rot type are well pronounced,
particularly for $\mu<0$ where both $R_N^{-1}$ and $I_c$ are strongly
suppressed, but most of the data lie below the lower bound given by Eq.\
(\ref{icrnpdiff}) for the pseudodiffusive limit, showing that the suppression
of $I_c$ is stronger than for $R_N^{-1}$. 
This observation coincides with another one for $m=2$ and $\mu>0$,
where the product $I_cR_N$ does not exceed 
the upper bound given in Eq.\ (\ref{icrnsubsh}) for the sub-Sharvin limit.
Both findings show that the reduction of $I_c$ in the tight-binding
simulation, when compared to the continuous model, which clearly demonstrates
the sub-Sharvin-to-ballistic crossover for $\mu>0$ 
(see Fig.\ \ref{icrnsmax10pan}), 
is more significant than for the case of $R_N^{-1}$ illustrated in
Fig.\ \ref{gicdiag700x240}(a).

Finally, in Figs.\ \ref{gicdiag700x240}(d,e), we present the critical current
--- skewness diagrams for the tight-binding results obtained for all
values of $m=2,4,\dots,128$ and $m=\infty$ considered in our study.
The values of $\mu/t_0$ are fixed at $5\cdot{}10^{-4}$ (for technical reasons,
we took a~small nonzero value), $0.1$ and $0.2$ in (d), or
$\mu/t_0=-0,1$, $-0.2$ in (e).
  In order to reduce the influence of Fabry-P\'{e}rot oscillations,
  each of the datapoints for $\mu/t_0=\pm{}0.1$, $\pm{}0.2$ represents
  the average (at a~given $m$) over energy interval of
  $\mu-\Delta{}\mu\dots{}\mu+\Delta{}\mu$, with $\Delta\mu=0.05\,t_0$. 

In contrast to the analogous diagram for the continuous model, see Fig.\ 
\ref{diagIcSmpots}, the datapoints for $\mu\approx{}0$ [grey circles] 
are now spread over a~much wider range, showing the tunneling-to-ballistic
crossover, and closely follow the single-mode approximation [solid lines].
This can be attributed to the presence of edge states propagating along the
free boundaries of the sample \cite{Wim09}.
The propagation through such states, albeit suppressed for flat potential
profiles ($m\rightarrow{}\infty$) due to the potential variation on edge
atoms, reappears for smaller $m$-s, as the weakly-doped area gets shrunk
to the narrow neighborhood of $r\approx{}r_c$.

The datapoints for $\mu>0$, as well as for $\mu<0$, follow the multimode
toy model, see Eqs.\ (\ref{tttoy}) and (\ref{tttoyicrn}) [dashed lines].
For $\mu>0$, the results stay within (or are very close to) the borders
of MDJT range, although previously (i.e., for the continuous model)  
the sub-Sharvin-to-ballistic crossover was observed. 
For $\mu<0$, the datasets are shifted such that they partly cover the MDJT
range, also taking the values below the lower bounds for $I_cR_N$ and $S$. 

It can be seen from Figs.\ \ref{gicdiag700x240}(d,e) that ---
except from the vicinity of $\mu=0$ --- 
a~generic reduction of transmission probabilities in the tight-binding model 
(if compared to the continuous model) 
leads to smaller values of both $I_cR_N$ and $S$, 
shifting the system characteristic towards the tunneling regime.

\section{Concluding remarks}
\label{conclu}

We have investigated the characteristics of the Corbino-Josephson setup
in graphene, including the normal-state conductance, critical current,
and the skewness of the current phase relation, supposing that the radial
profile of electrostatic potential barrier is tuned from a~parabolic to
a~rectangular shape.
The detailed interpretation of the barrier smoothing is left out of
the scope of this work; in principle, it can be regarded as an intrinsic
feature of of the device, grasped within a~self-consistent solution 
including the carrier diffusion and manifesting itself primarily by
the asymmetry of the conductance spectrum \cite{Pet14,Kam21,Kum22},
but one can also expect --- in particular, 
for ultraclean graphene-on-hBN devices \cite{Zen19} --- that additional
gate electrodes may allow one to tune the barrier electrostatically 
(at least to some degree). 

Our results show that the system behavior possible for the rectangular barrier
of an infinite height, earlier discussed in Ref.\ \cite{Abd18}, is
substantially enriched when smooth potentials are under consideration.
Namely, the product of critical current and normal-state resistance
($I_cR_N$) and skewness of the current-phase relation ($S$) analyzed within
the continuous model utilizing the Dirac-Bogoliubov-De-Gennes equation
exhibit --- when tuning the barrier profile --- crossovers between
three types of behavior possible for a~generic Josephson junction. 
A~standard Josephson tunneling (SJT), with the supercurrent governed by
a~single mode, is approached near the charge-neutrality
point for the values of the disk radii ratio starting from $r_2/r_1=4\div{}5$
and flat (i.e., nearly-rectangular) barriers,
graphene-specific multimode Dirac-Josephson tunneling (MDJT) occurs for the
tripolar doping and wide collection of barrier shapes, or for unipolar
doping and flat barriers, the ballistic Josephson effect (BJE) is
reconstructed for unipolar doping and smooth (nearly parabolic) barriers. 
Therefore, earlier findings of Ref.\ \cite{Ryc26a} are complemented by
demonstrating crossovers from SJT to BJE (unipolar dopings) and form
SJT to MDJT (tripolar dopings), both missing in the rectangular geometry. 

The picture following from the continuous model, and for the mesoscopic
device of approx.\ $0.5\,\mu$m in diameter, is confronted with the
tight-binding simulation of a~smaller, approx.\ $100\,$nm in diameter,
half-disk device. 
The relevant transport characteristics, for both the normal and
superconducting contacts, are suppressed due to finite-size (and
lattice-specific) factors, and the suppression is noticeably stronger for
superconducting contacts resulting in reduced values of the product
$I_cR_N$ and $S$ for nearly all physical regimes; however, the suppression
is marginal for SJT and MDJT regimes for flat (or nearly flat) barriers. 
These findings support the key observations for the continuous model,
suggesting that if the system size is within the mesoscopic range, the 
factors such as lattice discretization and trigonal warping of the dispersion
relation would not significantly affect the measurable quantities.
(Probably, a~better quantitative agreement can be achieved by using 
the method truncating the wavefunction within orthogonal polynomials, 
such as that recently implemented in the KITE software \cite{Joa20}. 

The Corbino-Josephson device in graphene, here discussed assuming the
short-junction limit and in the absence of magnetic field, are 
put forward as a~versatile system allowing not only for electrostatic
control over the critical current, but also for electrostatic switching
between different classes of Josephson effects in the mesoscopic limit. 
We hope this study would stimulate follow-up discussions, possibly
addressing the problem of macroscopic quantum effects 
\cite{Gua15,Mar85,Dev85} and quantum information \cite{Wen17} as the
(controlled) crossover between single- and multiple-mode Josephson effects
may also allow for temperature-independent control over the quantum
coherence.

\section*{Acknowledgments}
The work was partly completed during a sabbatical granted by the Jagiellonian
University in the summer semester of 2024/25. 
We gratefully acknowledge Polish supercomputing infrastructure PLGrid
(HPC Center: ACK Cyfronet AGH) for providing computer facilities and support
within computational grants Nos.\
PLG/2025/018544 (partly) and PLG/2025/018379.

\appendix

\section{The current-phase relation and charge-trasfer cumulants}
\label{apptns}

\begin{figure*}[t]
\includegraphics[width=\linewidth]{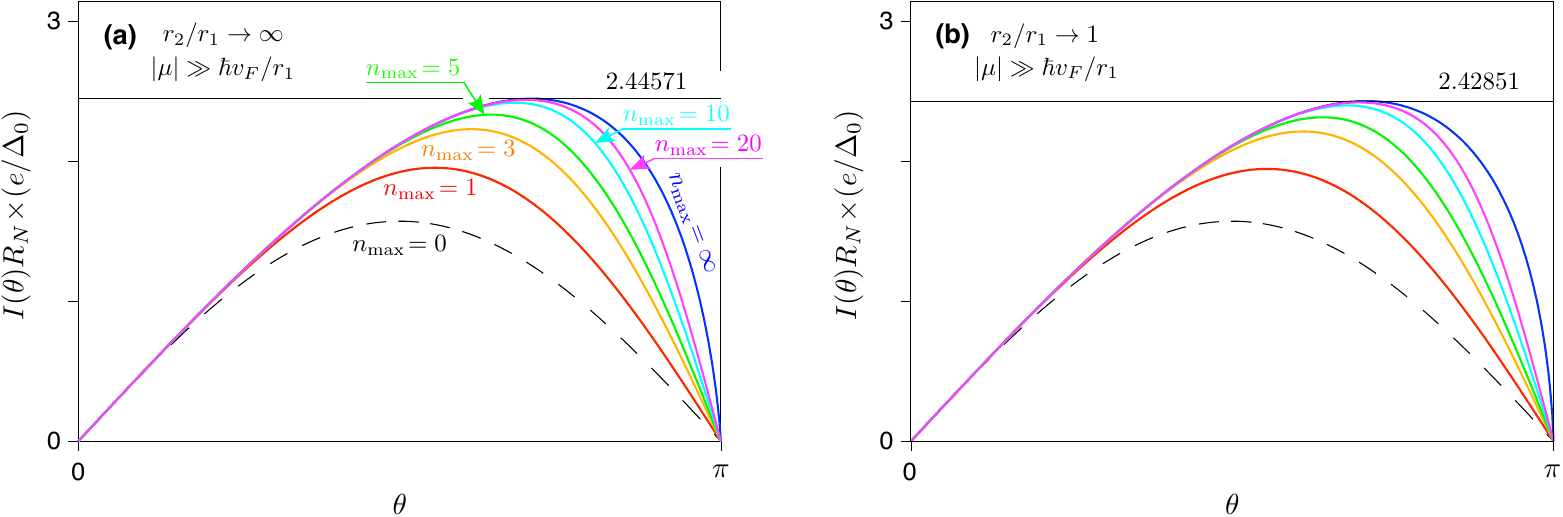}
\caption{ \label{ithdisk12ser} 
(a)--(b) Current-phase relation relation obtained by summing the terms
in Eq.\ (\ref{icrnseries}) up to $n=n_{\rm }$, specified for each line in (a), 
assuming the high doping ($|\mu|\gg{}\hbar{}v_F/r_1$) and 
(a) the wide-disk limit ($r_2\gg{}r_1$), or (b) the narrow-disk limit
$r_2/r_1\rightarrow{}1$. 
The case of $n_{\rm max}=0$ (black dashed  lines) reproduces the standard
Josephson tunneling (SJT), see Eq.\ (\ref{ijotunn}), for both limits. 
For the opposite, $n_{\rm max}=\infty$ case (solid blue lines) we perform 
the numerical integration in Eq.\ (\ref{ithsubsh}) (i.e., these data are
same as in Fig.\ \ref{ithdisk12}).
Since $I(2\pi-\theta)=-I(\theta)$, the presentation is limited to
the half of the period, $0\leqslant{}\theta\leqslant{}\pi$, this time.
The line-color encoding in (b) is same as in (a). 
}
\end{figure*}

In this Appendix, we point out that the current-phase relation $I(\theta)$
for a~superconductor-graphene-superconductor junction in the high-doping
limit, given by Eq.\ (\ref{ithsubsh}) in the main text, 
can be expressed via the charge-transfer cumulants in the normal state.
Specifically, expanding $(\sqrt{1-x})^{-1}$ with
$x=T_{j,\phi}\sin^2(\theta/2)$ and multiplying by the normal-state resistance
$R_N$ (\ref{ggdiskicoh}) yields
\begin{equation}
\label{icrnseries}
  I(\theta)R_N= \frac{\pi{}\Delta_0}{2e}\left[
    f_0(\theta) + \sum_{n=1}^{\infty} \frac{\langle{}T^{n+1}\rangle_{\rm incoh}}{\langle{}T\rangle_{\rm incoh}} f_n(\theta)
  \right], 
\end{equation}
where
\begin{align}
  \langle{}T^{m}\rangle_{\rm incoh} &= \frac{1}{2j_{\rm max}}
  \int_{-j_{\rm max}}^{j_{\rm max}} dj \frac{1}{2\pi}\int_{-\pi}^{\pi}
  d\phi\left(T_{j,\phi}\right)^m, \label{ttnincoh} \\
  f_0(\theta) &= \sin\theta, \\
  f_n(\theta) &=\frac{1}{4^n}{2n\choose{n}}\sin^{2n}(\theta/2)\sin\theta
  \ \ \ \ (n\geqslant{}1).  
\end{align}

Substituting $T_{j,\phi}$ given by Eq.\ (\ref{ttjdatta}) into Eq.\
(\ref{ttnincoh}) and considering the narrow-disk limit
($r_2/r_1\rightarrow{}1$), the resulting expression can be simplified to
\cite{Ryc25}
\begin{align}
  \langle{}T^{m}\rangle_{\rm incoh} &= \int_0^{1}d\eta
  \frac{2^{m-1}\Gamma(m-\frac{1}{2})}{\sqrt{\pi}\,\Gamma(m)}
  \left(1-\eta^2\right)^{\frac{m}{2}}z^{\frac{1-m}{2}} \nonumber\\
  &\times
  {}_2F_1\left(1-\frac{m}{2},\frac{1-m}{2};\frac{3}{2}-m;z\right),
  \label{ttmincohsubsh} \\
  \text{with } & z=\frac{1-\eta^2}{(1-\frac{1}{2}\eta^2)^2}, \nonumber 
\end{align}
where $\eta=j/j_{\rm max}$ and ${}_2F_1(\alpha,\beta;\gamma;z)$ is the
hypergeometric function \cite{Abr65}. 
(For a positive integer $m$, $\alpha=1-\frac{m}{2}$ or $\beta=\frac{1-m}{2}$
is a~non-positive integer; as a result, the function reduces to a~polynomial
of $z$.) 
For instance, the first four values are 
\begin{gather}
  \langle{}T\rangle_{\rm incoh}=\frac{\pi}{4}, \ \ \ \
  \langle{}T^{2}\rangle_{\rm incoh}=\frac{2}{3}, \ \ \ \
  \langle{}T^{3}\rangle_{\rm incoh}=\frac{3\pi}{16}, \nonumber\\
  \langle{}T^{4}\rangle_{\rm incoh}=\frac{8}{15}. 
\end{gather}

Similarly, in the wide-disk limit ($r_2\gg{}r_1$), Eq.\ (\ref{ttnincoh})
reduces to
\begin{align}
  \langle{}T^{m}\rangle_{\rm incoh} &=
  \frac{\sqrt{\pi}\,\Gamma(m+2)}{4\Gamma(m+\frac{5}{2})} \nonumber \\
  &\times 
  \left[
    2m + 3 - 2m\,{}_2F_1\left( \frac{1}{2}, 1; m+\frac{5}{2}; -1 \right)
  \right],  \label{ttmincohwide}
\end{align}
with the first four values equal to
\begin{gather}
  \langle{}T\rangle_{\rm incoh}=4-\pi, \ \ \ \
  \langle{}T^{2}\rangle_{\rm incoh}=\frac{40}{3}-4\pi, \\
  \langle{}T^{3}\rangle_{\rm incoh}=\frac{192}{5}-12\pi, \ \ \ \ 
  \langle{}T^{4}\rangle_{\rm incoh}=\frac{10624}{105}-32\pi. \nonumber 
\end{gather}

More generally, Eq.\ (\ref{icrnseries}) can be approximated using the
four lowest-order charge-transfer cumulants as follows
\begin{align}
  I(\theta)R_N\frac{2e}{\pi\Delta_0} &\approx
  f_0(\theta) + \left(1-F\right) f_1(\theta) \nonumber\\
  &+ \left(1 - \frac{3}{2}F + \frac{1}{2}R_3\right) f_2(\theta)
  \label{icrn4term} \\
  &+ \left(1 - \frac{11}{6}F + R_3 - \frac{1}{6}R_4\right) f_3(\theta),
  \nonumber 
\end{align}
where the $m$-th charge-transfer cumulant in the normal state is defined
via the electric charge $Q$, considered as a~random variable, that flows
through the system in a~short time interval $\Delta{}t$ and in the limit of
a~voltage $U\rightarrow{}0$ \cite{Sch07,Naz09}, namely 
\begin{equation}
  R_m = \frac{\left\langle{}(Q-\langle{}Q\rangle)^m\right\rangle}{\left\langle{}(Q-\langle{}Q\rangle)^m\right\rangle_{\rm Poisson}},  
\end{equation}
with the Poissonian value
$\left\langle{}(Q-\langle{}Q\rangle)^m\right\rangle_{\rm Poisson}=
e^{m-1}\langle{}Q\rangle$, and the average $\langle{}Q\rangle=\Delta{}t\,U/R_N$. 
(The Fano factor $F\equiv{}R_2$.) 
Therefore, representing the current-phase relation for superconducting
leads $I(\theta)$ as a~combination of linearly-independent functions
$f_0(\theta)$, $f_1(\theta)$, $\dots$, etc., one can estimate consecutive
charge  transfer cumulants for normal leads in the linear-response regime 
(at least in principle).

\begin{table}[!tb]
\caption{ \label{icrnseriestab}
The product $I_cR_N$ and the skewness $S$ obtained from the current-phase
relation given by the sum of terms with $n\leqslant{}n_{\rm max}$ in
Eq.\ (\ref{icrnseries}), with the cumulants given by Eq.\
(\ref{ttmincohsubsh}) for the narrow-disk limit ($r_2/r_1\rightarrow{}1$)
or by Eq.\ (\ref{ttmincohwide}) for the wide-disk limit ($r_2\gg{}r_1$).
$n_{\rm max}=3$ corresponds to the four-term
approximation, see Eq.\ (\ref{icrn4term}).
The values given in Eqs.\ (\ref{icrnsubsh}) and (\ref{icrnsubshwide}),
corresponding to the $n_{\rm max}\rightarrow{}\infty$ limit, are listed
in the last row for comparison. 
}
\begin{tabular}{c|cc|cc}
\hline\hline
$\,\ n_{\rm max}\ \,$ & $\,\ I_cR_Ne/\Delta_0\ \,$ & $S$
  & $\,\ I_cR_Ne/\Delta_0\ \,$ & $S$ \\
 & \multicolumn{2}{c|}{$r_2/r_1\rightarrow{}1$}
 & \multicolumn{2}{c}{$r_2\gg{}r_1$} \\ \hline 
$1$ & $\,\ 1.94411\ \,$ & $\,\ 0.108243\ \,$ & $\,\ 1.95247\ \,$
  & $\,\ 0.109920\ \,$ \\
$3$ & $2.21130$ & $0.220665$ & $2.22756$ & $0.223612$ \\
$5$ & $2.31262$ & $0.281842$ & $2.33142$ & $0.284759$ \\
$10$ & $2.39738$ & $0.358064$ & $2.41656$ & $0.359155$ \\
$20$ & $2.42566$ & $0.405813$ & $2.44331$ & $0.402860$ \\
$30$ & $2.42825$ & $0.414601$ & $2.44551$ & $0.410000$ \\ \hline
$\infty$ & $2.42851$ & $0.416008$ & $2.44571$ & $0.411023$ \\
\hline\hline
\end{tabular}
\end{table}

It must be noticed, however, that the convergence of the series in Eq.\
(\ref{icrnseries}), is --- in the case of graphene --- rather slow; 
for the illustration, see Fig.\ \ref{ithdisk12ser}. 
In particular, calculating the product $I_cR_N$ from the four-term
approximation given by Eq.\ (\ref{icrn4term}), one reproduces the exact values,
see Eqs.\ (\ref{icrnsubsh}) and (\ref{icrnsubshwide}), with $10\%$ accuracy. 
(To achieve $1\%$ accuracy, $11$ to $12$ terms are necessary.) 
More numerical examples are given in Table~\ref{icrnseriestab}. 

To conclude this Appendix, we point out that the approximation constructed
by taking only the first two terms in Eq.\ (\ref{icrn4term}), namely
\begin{equation}
\label{icrn2term}
  I(\theta)R_N\approx\frac{\pi{}\Delta_0}{2e}
  \sin\theta\left(1+\frac{1-F}{2}\sin^2\frac{\theta}{2}\right), 
\end{equation}
is insufficient for quantitative analysis except from a~small vicinity of
$F=1$; however, the above provides an intuitive explanation of the fact that
both $I_cR_N$ and $S$ grow simultaneously when the tunneling junction evolves
towards the ballistic parameter range.
During such evolution, the normal-state noise becomes sub-Poissonian,
with the Fano factor $F<1$, so the unharmonic component in
Eq.\ (\ref{icrn2term}) grows. 
More specifically, within the two-term approximation, 
varying $F$ from $1$ to $0$ shifts 
the pair $(I_cR_Ne/\Delta_0,S)$ from the point $(\pi/2,0)$
(corresponding to $F=1$) to the point $(2.0010,\,0.1192)$
(corresponding to $F=0$). 
The pseudodiffusive value of $F=1/3$ leads to
$I_cR_Ne/\Delta_0\approx{}1.8508$ and $S\approx{}0.0878$, surprisingly
closer to the $r_2/r_1=5$ values given in Table~\ref{icrntable} than to the
relevant values (for $r_2/r_1\rightarrow{}1$), see Eq.\ (\ref{icrnpdiff}),
yet still illustrating the qualitative trend. 

On the other hand, simultaneous growth of $I_cR_N$ and $S$ does not
necessarily appear for any continuous transition driven by the system
parameters, because the mesoscopic current-phase relation given by
Eq.\ (\ref{ijoth}) is complex enough to produce more cumbersome behaviors.
For instance, properly designed gate electrodes may, in principle, drive
the system from the narrow- to wide-disk limit (or {\em vice versa}) staying
in the high-doping regime. 
During such a~process, the values of $(I_cR_N,S)$ vary from those given in
Eq.\ (\ref{icrnsubsh}) towards those in Eq.\ (\ref{icrnsubshwide}); 
namely, $I_cR_N$ slightly increases, whereas $S$ slightly decreases,
following the trajectory that is approximately perpendicular to
the line representing the multimode toy-model, 
see Eq.\ (\ref{tttoyicrn}) and dashed line in Fig.\ \ref{diagIcSdisk}. 




\end{document}